# An Entropic Model for Assessing Avian Flight Formations

J.A. Sekhar
University of Cincinnati, Cincinnati, OH, 45221, USA, and
MHI Inc. Cincinnati, OH 45215, USA

**Abstract:**

Why do birds fly in well-defined formations?
The observed formations of birds-in-flight are studied with a thermal model. Avian formation-flying is tested as a self-organization process. The maximum entropy production rate (MEPR) postulate is examined to determine its applicability to the observed "V" formation during the long-distance migration of large birds. A brief discussion of the principles of the rate of entropy generation maximization or minimization is first presented. The "V" formation is compared to other closely related, but distinctly different formations that can also be adopted by a flock. The thermodynamic analysis from the results of the model indicates that the "V" formation (named the SV formation in the article) is found to be the pattern that optimizes energy use for the flight duration. This is accomplished with savings in the formation's thermal energy output (heat) by the trailing birds. The MEPR postulate is found to hold for the "V" formation. The formation also provides for the highest comfort temperatures, perhaps thereby allowing for high altitude (low drag) flight levels to be accessed by the birds. The results of the model point to thermal awareness as a possible key driver for organization into a "V" formation. The thermal model appears to validate *all* the major experimental observations of avian formation flying. Regardless, without detailed thermal measurements to validate, and without a physical basis for the constant used in equation (C4d), it is not yet definitive that this thermal model is valid for avian flight formation. The analysis appears to indicate that the organization into patterns of live objects may follow the MEPR principle for self-organization in a manner previously noted for chemical and metallurgical processes.

# 1. INTRODUCTION

Avian-flock self-arrangement into a "V" formation during long migratory flight passage is a commonly observed pattern for large-size birds like the Canada-geese and Cranes. The "V" formation is possibly a process of self-organization with an energy-optimizing objective [1-10]. This is the main hypothesis that is tested in this article.

Observations indicate that birds that fly in the "V" formation spend only half of the energy compared to when they fly alone or in some other formation [1,2,3,9]. Analyses of flock formations with time-lapse photographic techniques [9] have analyzed bird positions in the pattern and found them to be approximately located such that they gain energy from the work done by the leading birds. The lead bird (at the apex of the V formation) periodically trades places with birds further back. Over the long haul, all birds thus expend the same energy [1.2,9]. The temporal wing positions of the birds in flight appear to also have some coordination [9] when they fly in a "V" formation.

If the time scales for energy availability and partitioning into work and heat are of similar durations for long flights, the maximum endurance (time aloft) is obtained for flight conditions that correspond to the minimum rate of energy expenditure (J/S, Joules/second) by the birds for a fixed amount of fuel-energy available. A minimized rate of fuel expenditure, per kilometer flown, appears to be associated with the "V" formation flights. Birds apparently can conserve energy when required. For example, birds find methods of conserving energy, even if the flights must be made in hypoxia conditions [6].

Formation-flying is possibly a way of minimizing waste energy (the energy transferred between birds and the atmosphere but not converted to work). Previous models for the assessment of avian flight for such energy benefits by invoking formation flying [1,2,3,6,7,8,9,10] have mostly considered wing tip eddy forces as helping trailing birds perform less work by flying in the "V" formation. Although this is a distinct possibility, i.e., utilizing waste work directly by the trailing birds, the work-only models have not always been rigorously tested for different formations to qualify the "V" formation as the optimum one.

In such eddy models, the trailing birds adjust their relative height during flight to take advantage of the airflow patterns. Eddy generating work by a bird, or a flock must also be accompanied by a corresponding heat output which has seemingly not been adequately considered for an energy optimization calculation. In addition, it is known that the wings are not perfectly synchronized for trailing birds to ensure optimum use of work [9,10]. In the case of aircraft, wingtip eddies when they impact trailing aircraft, lead to considerable instabilities *i.e.,* vibration. This has not been reported for birds on long-haul flights.

There are two significant experimental observations that any model must satisfy, namely (i) the energy expenditure per bird should decrease with increasing flock size and (ii) the "V" formation flying is the preferred formation over other configurations (shown in Figures 1, 2 and 3). Similar in many ways to the "V" formation pattern, other patterns are also sometimes observed for avian flight. An example is the one-leg "V" formation called the staggered formation shown in Figures 1 and 2.

Formation flying appears to be self-taught, not a genetic hand-down information[8]. It is readily recognized that there may be other social benefits beyond energy-optimization for flying in a formation compared to individual flights. This article probes the energy exchange and entropy generation during formation flying. A thermal model for formation flying is explored. This is a simple Lagrangian model in contrast with the more detailed Newtonian models that are commonly employed to study formation flying [1,2,9]. The thermal model only considers the possible thermal influences on energy efficiency and pattern formation. It is not a replacement for the force-enhancing models that are in place. Unless detailed experimental data is available it is not possible to select the key parameters that are dominant.

The thermodynamic model in this article allows for testing various equally probably formations for energy efficiency. It is assumed that birds in a formation fly at the same level (altitude).

A few possible basic flight formations are shown in Figures 1 and 3. We first model the four formations shown in Figure 1 and then progress to larger flocks shown in Figures 2 and 3. These are labeled by their type and number of birds. Some like the SV (Figure 1d) formation ensure that all birds (except for the leader) experience the same environment regardless of the flock size. Figure 2 shows the various type of "V" formation flights that have been photographed and published [9]. Figure 3 shows possible formations for five or more birds. This article examines the energy expenditure rate and the entropy generation rate during pattern flying for testing the *Maximum Entropy Production Rate, (MEPR)* postulate for predicting the dominant patterns [4,5,7,8,9,11-22].

In this article, all birds are assumed to be of one size. Not all birds will be exactly of the same size. Some birds could be oversized. In some formations, the bird can slip out of formation without significant disruption to the other birds. In other formations (patterns) this could cause considerable disruption. For example, in the I and H formation, the symmetry of the formation is easily disturbed if a bird is oversized. An oversized bird is more easily accommodated in the SV formation as it offers more degrees of freedom for maintaining the operating thermal conditions even if a bird or two is oversized. In the SV or ST formation, the birds can also adjust their angle (relative to the wind) so that a horizontal oscillatory component of flight is recognized, yet the morphology of the formation is not significantly disturbed. This is another way in which the ST and SV formations are different from other formations.

Although in this article the reasons for formation flying are sought from energy and entropic principles, it should also be recognized that the "V" formation is a formation that allows full visibility to every bird while recognizing a lead bird for flock instructions. It additionally provides for easy rotation between positions for leaders and follower birds (Figure 1) without necessarily changing the flying environment when a switch occurs.

# 2. Entropy Generation Maximization or Minimization? The MEPR Postulate for Pattern Formation.

One of the key objectives of this article is to compare the rate of entropy generation between various possible flight formations. Patterns that are noted to form in solidification, wear, and other materials processes, have been found to follow the MEPR postulate [4,12-22]. The entropy rate maximization has been discussed as an important governing principle that guides patterns observed for galaxies, turbulent flow, and patterns observed in Hele-Shaw cell experiments [8,16]. Yet the applicability of the principle is not fully established. Towards this end, we first briefly review the principle of MEPR (Maximum Entropy Generation Rate). The MEPR postulate [4,8,12-20] compares the entropy generation between different pattern-formations to identify the pattern that maximizes the entropy generation per unit volume [4,12]. For an open-flow process, Reiss [7], Martyushev [8], and Veveakis, et.al., [13] have shown that the entropy generation is maximized for constant thermodynamic forces and is minimized for constant fluxes [7]. A brief review of MEPR is provided below.

Following the pioneering publications by Prigogine [23] and Zeigler [24], there have been competing articles regarding the use of either the postulate of entropy rate minimization, referred to as MinEP or PPEDMR [12,23,25,26,27,29,30] or the principle of entropy rate maximization, referred to as MaxEP, MEPP or MEPR [7,8,9,12, 24,25, 26,27,29,30] for describing reaction pathways or pattern formation. The conflicting use of either principle arises because the second derivative of the entropy generation rate often displays an inflection point when it is assessed. Several comprehensive and critical analyses and summaries are available that discuss the key issues surrounding the use of maximum or minimum entropy generation rates for an analysis [25,26,27,28]. A simple explanation for MEPR is offered below.

When heat is transferred between a hot and cold reservoir at temperatures $T_1$ and $T_0$ respectively at a rate equal to $q^{\cdot}{}_1$ and $q^{\cdot}{}_2$, a part of the energy can be converted to useful work ($w^{\cdot}{}_{\text{actual}}$). When not extracted as work, a certain amount of work-potential loss $T(\dot{\phi})$, occurs because the quality of energy is degraded. Here, T is a temperature between $T_1$ and $T_2$, and $(\dot{\phi})$ is the entropy generation rate. It should be noted that the traditional Carnot expression does not use a rate term in it. When time is not considered, the best efficiency can be thought to produce very low power because only then can it be reversible as it happens slowly (for the reversibility condition to hold).

An expression of the second law for qualifying the thermodynamic efficiency (heat to work at finite rates) can also be written in the following manner with expressions that use power instead of energy:

$$w^{\cdot}{}_{\text{actual}} + T\phi^{\cdot} = q^{\cdot}{}_1[T_1 - T_0]/T_1 \qquad (1)$$

The efficiency of the work production rate for external work is highest when the $T\phi^{\cdot}$ is zero (Carnot efficiency).

The energy gradient that is related to entropy generation, $T\phi^{\cdot}$, is used to transport fluxes. For example, the presence of a temperature gradient enables heat flow. Similarly, mass flow can occur with a pressure gradient. However, there is a limit to the amount of flux that can be transported (alternately thought of as a limit to the entropy transport) by any pattern whether a solid with a defined crystal structure or an atomic arrangement that defines the viscosity of a fluid. In the case of thermal transport in a solid, there is a lattice-imposed limiting condition from the thermal conductivity, particularly if the volume is constrained. To increase this rate when entropy generation is not able to be dissipated from the volume a new pattern could emerge [4]. This is easily visualized for a liquid or gas where turbulence or complex convective currents emerge to aid heat and mass transport when the entropy transport requirements become severe. In two-phase processes e.g., solidification, complex solid patterns called dendrites can emerge because the solids with finger-like dendrites can conduct heat better than other solid-liquid morphological variations of the intermixed two phases [4,12-20].

Thus, there is the possibility of new patterns forming for processes that include gradients of temperature, pressure, electrical, magnetic, or chemical potential. For such processes at a steady-state, entropy must be exported from the boundaries of a control volume. In such processes, there is work absorbed to create or maintain a pattern within a control volume to enable this export of entropy. The pattern may cause unusual defects to emerge that assist entropy transfer. For example, defects such as segregated dendrite boundaries are noted during the solidification (directional crystallization) of alloys [4,12].

As several competitive patterns are often feasible by spatially rearranging the core repetitive features, i.e., the building block for the pattern, there is a need to examine which principle(s) if any can predict a dominant pattern. It should be noted that the various possibilities in the pattern may require a change in the control volume dimensions [4]. Useful patterns must also display stability. Patterns can discontinuously change into different patterns when under duress [4]. The alternative is chaos. The amount of energy expenditure that is required by different patterns for their formation is not the same. This can often physically manifest in the differences in curvature, spacing, or types of other features that describe the pattern at various microscopic or macroscopic scales [4]. The energy expenditure for this work often comes from within the system. A classic example of pattern changes is noted when a fluid is heated from the bottom of a container.

Inside a body of fluid in a container that is heated from underneath, various patterns can be considered to *compete* for dominance depending on the driving force (temperature gradient) and other body forces. There is an ordered pattern of molecules that contributes to conduction heat transfer. There is also a larger-scale fluid flow pattern from convection which is also possible for enhancing the heat transfer. If the heat conduction (thermal conductivity of the fluid) is high, the ordered pattern of molecules i.e., from the short-range ordering of molecules in the liquids, is adequate for energy and entropy transport. On the other hand, if the heat conduction is very low, a large temperature gradient may be formed, and convective-cell patterns will dominate (generally driven by gravity or surface tension). For liquids, the competition between conduction and convection is tracked with the Rayleigh number (Ra). When Ra is>~1000 (in earth gravity conditions) for liquids, it leads to the convection pattern showing dominance. If the Rayleigh number becomes very high, a new pattern, called turbulence, can replace the convective cells. Similarly, during solidification from a liquid state, the competition between metallic glass and ordered crystalline-lattice formation, i.e., different patterns have been noted to be influenced by the Biot Number (Bi) [4].

For a spontaneous process (such as heat flow from a high temperature reservoir to a low temperature reservoir), when there is no objective for obtaining useful external work, Equation 1, can be recast as:

$$\dot{q}\ [T_1-T_0]/T_1- T\dot{\phi} =0 \qquad (2)$$

$$\dot{q}= T_1.T.\dot{\phi}/(T_1-T_0) \qquad (3)$$

The grouping, $(T\dot{\phi})$ is the rate of **external** work-potential loss. This lost work potential enables (i) the flux, which is a conjugate term in the Onsager force-flux relationship [31], and (ii) provides the energy for *pattern-formation and if required for its maintenance*. Effectively this implies that if there is an expected “natural” tendency for maximizing the rate of heat transfer, then the term $T\dot{\phi}$ should be maximized and vice versa. There is no formal law that mandates such a rate maximization, except as noted in the principle of least action [32,33] for gravity and other potentials [32,33]. Maximizing the heat transfer rate in Equation (3) is the same as minimizing the resistance to heat flow. This can be enabled with a change in the patterns that impact heat transfer.

There are several reasons why a principle like the least action postulate, may also hold for thermal and chemical potentials and in turn influence pattern formation. It was shown previously [12,19] that as the driving force decreases, older patterns can be recovered which implies that the process conditions influence the pattern (morphology). For stable thermal, biochemical, and chemical processes, there is an expectation of a steady state which often is approached asymptotically [16]. An unstable system diverges from the steady-state. As a dynamic, open system is expected to approach a stable steady state (which is the open system equivalent of equilibrium for a static, isolated, closed system), the rate of entropy production must be maximized to maximize the heat transfer rate and the entropy export rate [4]. Thus, without a high degree of mathematical rigor, it can be argued that the maximum entropy generation rate (per unit volume) [4,7-21] is demanded by the conditions required for establishing a steady state for a given volume. The postulate of MEPR compares the entropy generation rate for various patterns that can form at any scale during a process. MEPR is thus like the principle of least action [32,33].

# 3. AVIAN FLIGHT MODEL.

The four formations (patterns) shown in Figure 1 are for geese. Each bird is assumed to occupy a volume defined by a cube of length L (Figure 1 (a)). The formations that are shown in Figure 1 are the Horizontal, H in Figure 1(b); In-line, I in Figure 1(c); Separated V or SV, in Figure 1(d); and the Staggered ST, in Figure 1(e).

These formations are compared for energy conversion efficiency and entropy production rate with the model described below. In the model, the heat is removed from the control volume only by the flow of the air (relative wind) opposite to the flight direction. There is no heat loss perpendicular to the flight direction.

Although the flight speed of birds is determined by a variety of factors, the size of the flock has a significant impact on how fast the birds can fly when in a formation. It has been recorded that less energy is expended over time, for the same flight distance, as the flock size increases [1,2,3]. Any model that describes formation flying must predict this lowering of energy expenditure with increasing flock size.

The expenditure of energy during the flight of a bird involves both work and heat exchanges, and heat production (by the bird). The work is done by the bird for providing lift and thrust. The heat warms the bird, excess heat is jettisoned as required by the second law when converting heat to work. In the model developed below, the heat transfer rate is assumed to be rapid which is also the experimental report [36]. The heat travels (leaves the control volume) with the relative wind.

The symbol $\dot{e}$ is the power expended by each bird. The rate of energy change of a bird inside a control volume is $\dot{e} = \dot{q} - \dot{w}$ where $\dot{w}$ is the work done by the bird and $\dot{q}$ is the heat exchanged. Useful energy for the work and heat exchange is obtained from the metabolic process. The chemical free energy partly transfers to work, and the rest to heat. The heat produced is used for warming the bird and is also transferred (exchanged) with the flowing air when the bird is in motion. Note that $\dot{q}$ only refers to the amount of heat exchanged with the flowing air, however over the long haul, a lower $\dot{q}$ signifies a lower energy production by the bird. For a steady-state condition, the mass of air entering the control volume is equal to the mass of air leaving the control volume. A key assumption made in the model is that the heat produced is well distributed quickly in the control volume along the bird's flight direction but is not shared in the x-direction. Thus, in the trailing control volumes, there is a temperature distribution along the x-axis that is recognized at the exit plate as shown in Figures 1 and 3. The thermal energy leaves the control volume only in a direction opposite to the flight direction and only with the wind. Thus, a change (savings) in the $\dot{e}$ signifies energy savings by the bird. In the model, the heat is removed from the control volume only by the flow of the air i.e., there is no heat loss perpendicular to the flight direction. There is significant evidence that indicates that birds are adept at extremely rapid regulation of their temperatures by interacting with the environment [9,10 34-42]. In this article heat transfer and heat production only are of importance for understanding efficiency. These are expected to be controlled by the environment temperature. It should be recognized that productivity and the ability to do work, are both influenced by external temperature, but they signify different outcomes. Productivity is a behavioral aspect whereas the ability to perform work, is a thermodynamic term. Strong variations in the behavioral patterns are recognized for several birds when the environment temperatures vary over ~20-30K [34].

Work is performed by birds by wing movement. This work cannot be produced without some heat being produced and jettisoned [10,36]. The amount jettisoned may not be small in relation to the work terms. Energy optimization can arise from minimizing the wasted heat output. This is a recognized issue even for humans when converting chemical-free energy from ingested food for work production. Humans typically are only ~25% efficient in converting energy to work, the rest being dissipated as heat. The model relies on experiencing and responding to small temperatures changes by the bird. A change in temperature impacts fatigue and as will be noted in the results below, impacts energy optimization. It appears that the outside temperature is a key parameter for a bird during its flight to adjust its wing morphology. The role of the feather coat in decreasing the temperature gradient between the body surface and the environment is of extreme importance concerning thermoregulation [36].

As the bird soars, she can adjust her wing-tip patterns (somewhat like adjusting airplane flaps and slats) and the angle of attack. For a bird, the core region (inner regions of the bird) is maintained by the shell (outer regions) which can quickly exchange heat with the environment. The core temperature is regulated by the metabolic processes *and* the heat production in the skeletal system. The core temperature is maintained at a relatively constant level, thereby providing constant temperature to the central nervous system, visceral organs, and even to a part of the skeletal musculature. The core region is where heat production occurs and maintains the physiological processes vital to the life of the bird. The shell region includes the remaining skeletal musculature, the skin, the feathers, the uninsulated wing extremities, and some fat [35,36]. It is known that for wind velocities in the 2-5 m/s, range birds can alter their temperature easily by ~5K during a day while interacting with the environment [36].

Starlings execute a very coordinated pattern to create warmth. Up to 4K higher than the resting temperature in starlings have been recorded [10]. Sometimes the skin temperature on the breast is within a few degrees of even the core temperature [10]. Insulation, which is related to the wing and feather morphology, is adjusted to maintain temperature [10,35,36]. This changes the amount of work proportioned between lift and thrust. Any change in the wing profile also changes thermal regulation [36].

One of the key assumptions made in this article is that birds can rapidly respond to an external change in temperature. There is some experimental evidence available that supports this assumption. Torre-Bueno [10] has shown with experimental measurements that all heat that flows through the skin at a given point must also flow through the feathers. The feathers interact with the environment. Veghte et. al. [41] have shown that for birds, even when at rest, the face and bill are warmer than the outer surface of the feathers, at least for the lower environment temperatures. It appears that the morphological features of the exterior of the birds can rapidly exchange heat with the environment, indicating that they can lose heat rapidly at the feather tips. The thermal resistance of both the skin and feathers can be adjusted, the former by vasodilation and constriction, and the latter by feather erection [41]. It has been noted that the skin temperature is only a weak function of the ambient, in fact, even falling in several bird skin locations when the ambient temperature increases. *Thus, it appears that a bird can adjust the heat production quickly to respond to the outside temperature.*

It is recognized that the wing mechanics, internal heat production or heat distribution between skin and core, *and* the fluid dynamics of bird flight along with the heat dissipation environment are quite complex to comprehensively model in detail. Yet, regardless of such complexities, an energy and entropy balance averaged over a large duration flight can be attempted with a simple model by assuming that the bird velocity and exit air velocity from the control volume are the same. The expenditure of energy may involve use of discontinuous use of stored energy or undigested food (birds like geese store food in an area region of the bird called the Crop) as well as draw on heat production from several metabolic processes. To minimize the energy-cost of temperature regulation for thermoregulation, birds use a variety of orientational and morphological variations, along with changes in behavior to adjust the rate of heat loss or heat gain [38-42]. Thus, without working through the complexities of the forces at play, it may be possible to calculate the benefits of formation flying from a thermodynamic model.

In this article, we compare the energy use at a fixed velocity of flight and fixed work done. The drag force thus only peripherally enters the calculations. For completeness, we note that the drag is composed of two parts, (1) a part termed *parasitic drag* that increases with the square of the flight velocity, and a part called *induced drag*, or drag due to lift, that decreases in proportion to the inverse of the flight velocity squared [3,11]. For overcoming the drag, the force balance with velocity takes the form $C_1V^2+(C_2/V^2)$ [3,11], where the $C_1$ and $C_2$ are constants and V is the flight velocity. In previous models [1,2.3,10] the vortices created by drag associated with lift (that are normally undesirable for airplanes because they create a downwash that increases the induced drag on a wing in flight) provide a benefit for *avian flight,* as it is also accompanied by an upwash that can be beneficial to a trailing wing flying behind and slightly above the first to gain free lift. The model considered herein assumes t in the power curve vs velocity for avian flight is at the minimum drag configuration or at least follows the latter part of the traditional curve [11] where power and velocity are positively correlated. If W is the thrust (in Newtons), the rate of work produced is $\dot{w}$ = W. V {i.e., the work rate to overcome drag) + (work-rate to maintain lift)}. At constant forward velocity and altitude, W is equal to the drag force during steady-state flight. For aircraft flight, and by extrapolation, for avian flight, the time-aloft (endurance) is maximized for a fixed quantity of available energy (the fuel) when the rate of energy usage is lowered. To maximize the range, it is necessary to minimize drag for a given weight. In this article, the energy benefit is correlated to the reduced amount of heat produced. The drag, per se, is therefore not important for the objectives of this article, except to note that the total power and the total entropy generation are a function with a $\sim V^3$ dependency. From the calculations presented below, we seek the lowest power expended for a given velocity of flight, as the measure of efficiency.

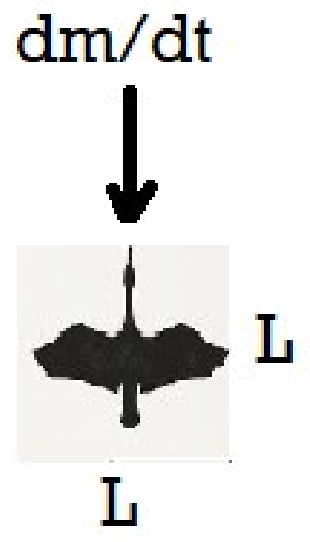

(a) Individual lone bird flight (LB). The sweep area per bird is $L^2$.

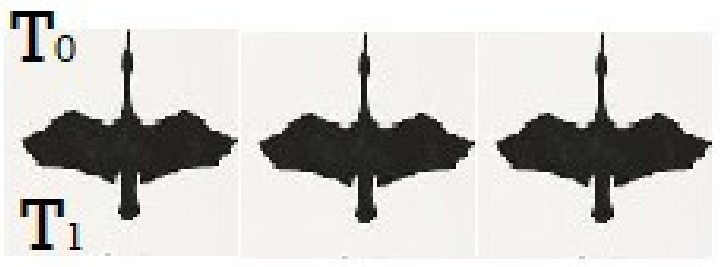

(b) Horizontal (H) Formation. The sweep area per bird is $L^2$ for the H3 formation

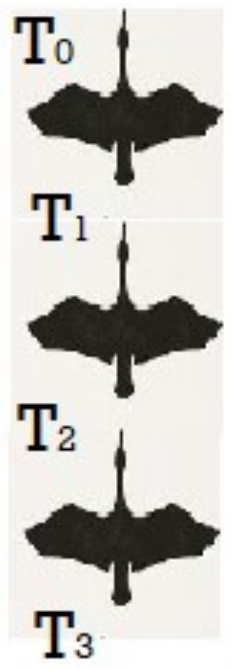

(c) In-line (I) Formation. The sweep area per bird is $L^2/3$ for I3 formation.

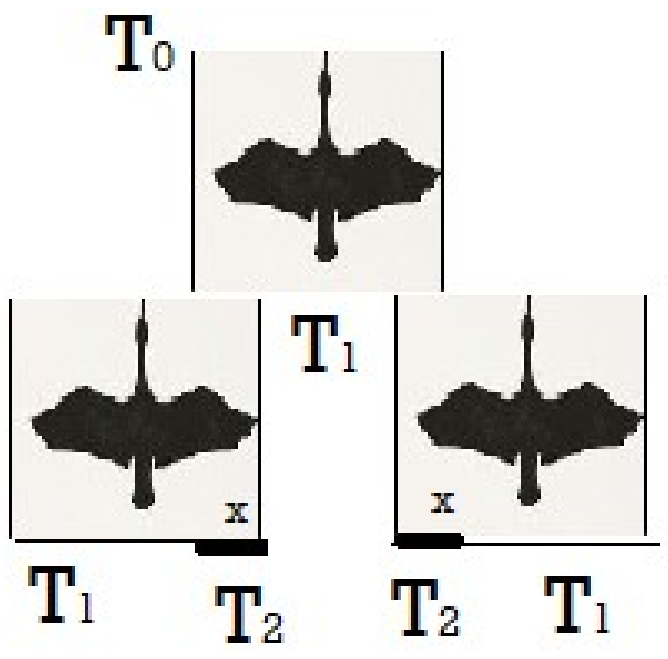

(d) Separated V-shape (SV) Formation. The sweep area per bird for SV3 formation at x/L=0.5 is $2/3L^2$

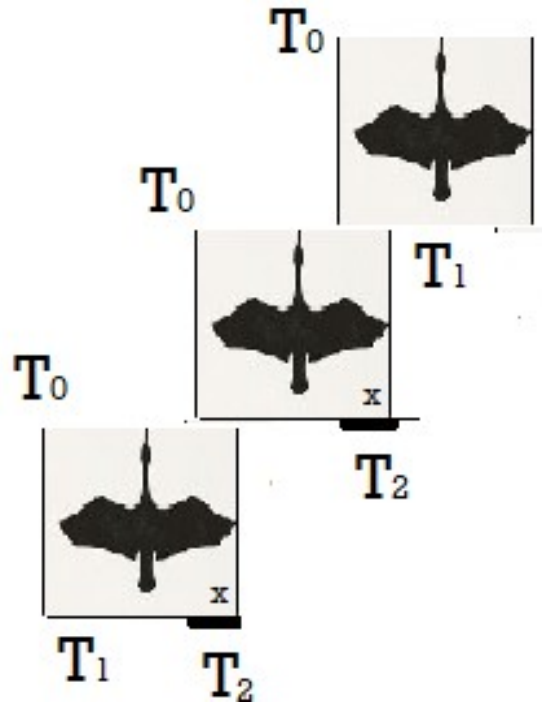

(e) Staggered Formation (ST) Formation. The sweep area per bird for the ST3 formation at x/L=0.5 is 2/3 $L^3$

**Figure 1**. **(a)** is the individual bird, **(b), (c), (d), and (e)** are for formation flight morphologies with marked exit temperatures.

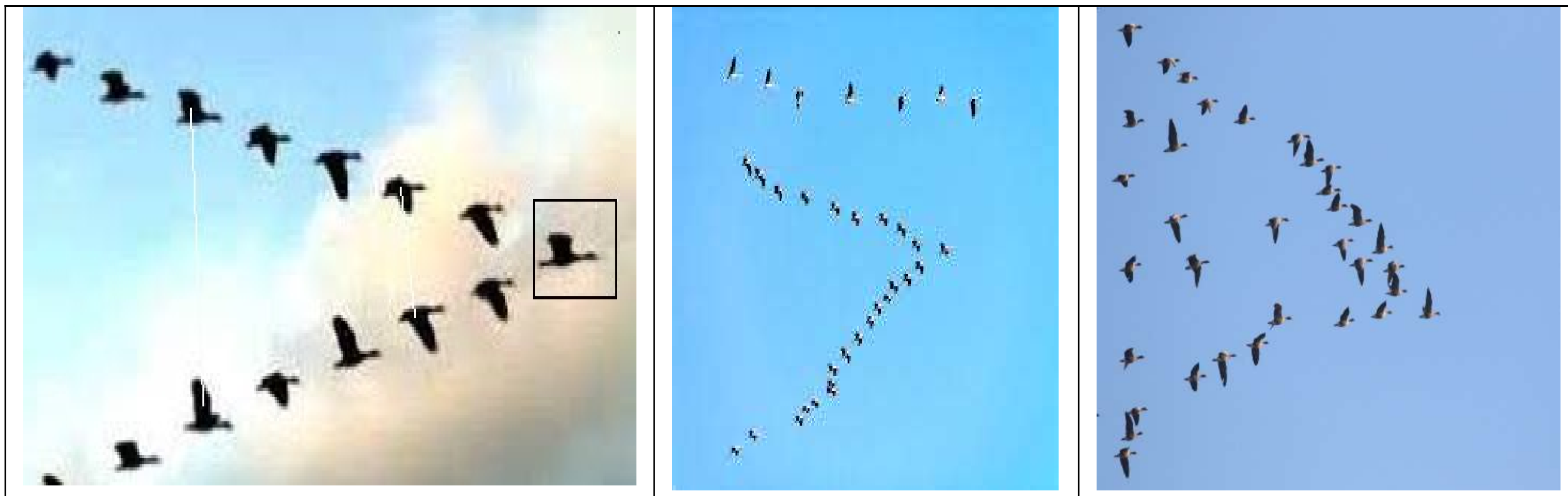

(a) (b) (c)

**Figure 2**. **(a)** “V” formation of flying geese. The flight path is left to right. Notice that the wing positions are somewhat synchronized but sometimes out of phase at any of the equivalent bird positions behind the leader. The birds fly in a separated V formation with the separation partner distance increasing with levels behind the leader. Original Picture Credit before modification: http://www.strategistblog.com/2020/10/flying-in-remote-formation.html. The control volume of the lead bird is shown.
**(b)** Geese also fly in a ½ V formation (or staggered formation). The picture shows a staggered formation line of birds flying alongside a “V” formation. In this picture, there could be two leader birds, one for each formation or just one leader trading places when required with the second formation. Credit https://en.wikipedia.org/wiki/V_formation.
**(c)** Geese also fly in a filled or tight V formation. Credit https://www.holkham.co.uk/blog/post/pinkfeet-horizons.

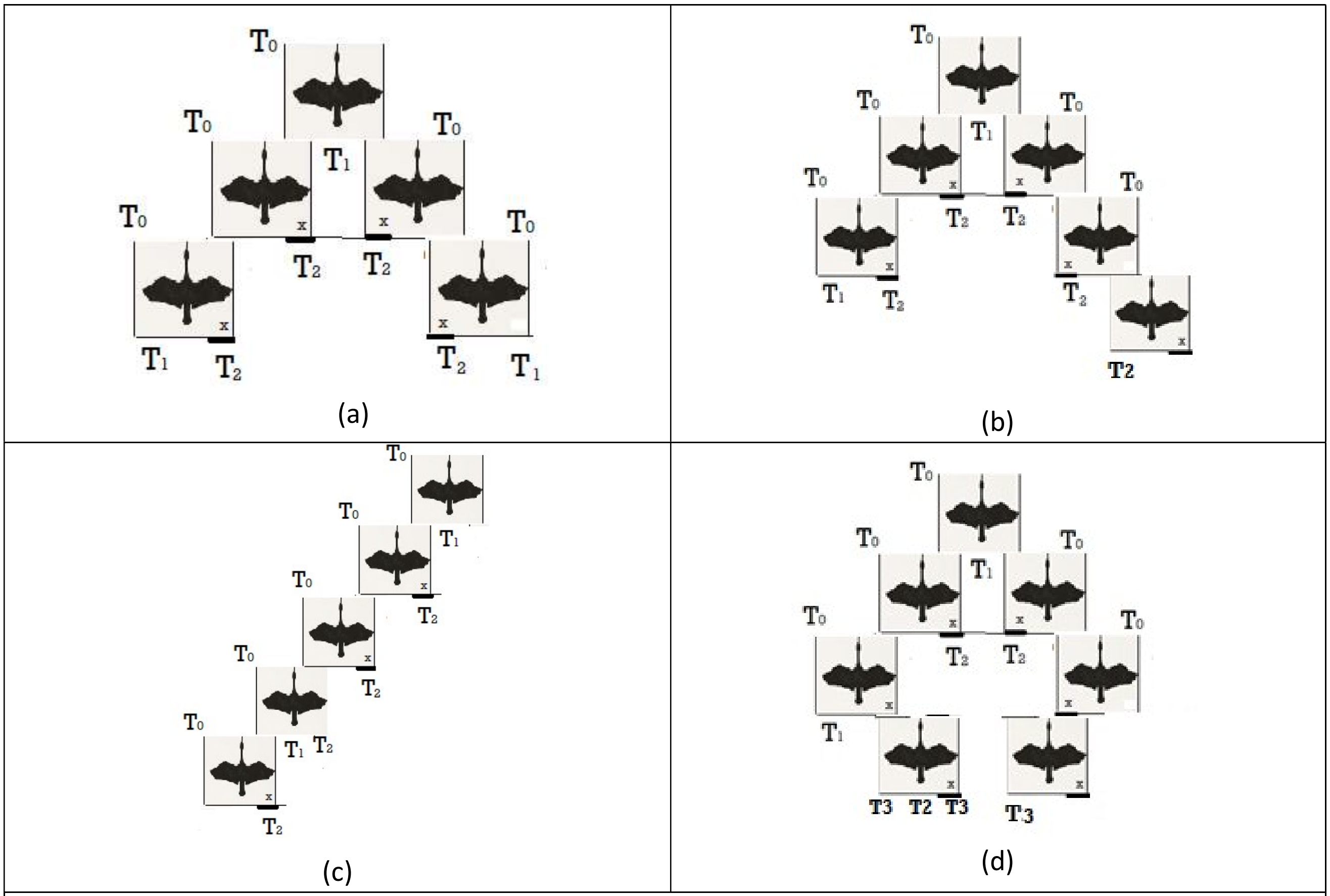

**Figure 3**. Five and six bird formations of the SV and ST patterns. **(a)** 5- Bird V Formation (SV5), compare with Figure 2(a) for a lesser number in the flock. **(b)** 6- Bird V Formation (SV6), compare with Figure 2(a). **(c)** 5- Bird Staggered In-line Formation (ST5) compare with Figure 2(b), **(d)** 7- Bird Tight V Formation (SVT7), compare with Figure 2(c).

The exit-plane temperatures are subscripted by the number of birds and the type of formation as illustrated in Figures 1 and 3. The frame of reference can be attached to the bird in which case the control volume encounters mass in and out of the air(gas) which crosses the control volume boundary in front of the bird and leaves from behind the bird. (Volume $V=L^3$). Here $\dot{m} = L^2.\rho.$ is the steady −steady-statevelocity of the bird, ρ is the air density. For the H and I formations, the exit plane is assumed to have a fixed temperature. For the SV, ST, and other related V formations (see Figure 3) the exit plane for a bird cell (control-volume) can have a variable temperature that depends on the location. The rate of heat production, $q^{\cdot}$, at a steady-state, along with the heat exchange establishes the temperature (T) of the bird and her control volume. Each bird in the flight formation is assumed to produce the same amount of work per unit time regardless of its position. To gain efficiency from formation flying, the extra heat that enters the bird control volume from her leader is converted to work (or less heat is produced by trailing birds as the environment is warmer). In thermodynamic terms both the situations are equivalent.

If the work expended by a bird is the same to fly at a fixed velocity at a given altitude, the improved efficiency will result if a lesser amount of heat is produced (when flying in the most efficient pattern). Consequently, the exit cell temperature difference can become lower with receding levels e.g., ($T_2$-$T_1$) is smaller than ($T_1$-$T_0$). However, it is possible but perhaps unlikely that during the entire migratory journey, each trailing bird becomes hotter in a formation because not all the extra heat (from her leader(s)) is convertible to work or harnessed to lower heat production in the trailing birds. Noting that birds exchange places, we shall assume that all trailing birds experience the same average temperature during migration. In addition, certain geometries of patterns related to the "V' formation are better able to allow a common temperature for all birds except the leader. *For example, although a continuous increase in temperature can happen in the I formation with every receding level of the trailing bird, the SV and ST formations provide a configuration where the trailing birds are all at the same temperature as shown in Figure 1.* Consequently, for all such patterns even without an exact temperature being known, one can identify the pattern that yields the lowest bird temperature for equivalent placement of the birds in that pattern and thereby predict the most optimal pattern for energy optimization.

At steady-state conditions, the entropy rate change of a control volume is zero [4]. The entropy generation $\phi^{\cdot}$ is the rate of entropy generation in the total number of cubes (control volumes) of a formation (Figure 1 and Figure 3). In the equations developed below, $C_v$ and $C_p$ are the specific heat at constant volume and constant pressure respectively, for air. The symbols $T_0$, $T_1$, $T_2$, and $P_0$, $P_1$, and $P_2$ are the temperatures and pressures at locations of entry, and subsequent levels. We first test the model for three bird formations and five bird formations as it is now well understood that this is the size of the flock where the maximum benefits are recognized for drag [1,3]. Each case below represents a particular formation. The energy and entropy balances are made for an open system i.e., the PV (pressure multiplied by volume) energy is captured in the enthalpy. The x (horizontal), y (direction of flight), and z (vertical) directions are centered in the control volume. Nine types of formations or flock sizes labeled Case 1 to Case 9 are developed below for the assessment of energy and entropy rate balances. The dependence on the exact position in a particular formation is possible by varying x/L. Note that x/L is a fraction (0 to 0.5) for the overlap (see Figures 1 and 3). We note that x/L cannot be zero which is a condition that precludes any interaction between leading and trailing birds in the model.

**Case 1: Single Bird, (also referred to as Lone Bird LB).**

Assume that the control volume (CV) with energy $E_{CV}$ is at a steady state. $T_1$ and $T_0$, and $P_1$ and $P_0$ are the temperature and pressure at locations 1 and 2. $D_{KE}$ is the rate of change in kinetic energy. The energy balance at steady state yields:

$dE_{CV}/dt = \Delta_{KE} + [e^{\cdot}] - m^{\cdot}C_p.(T_1-T_0) - d[V\Delta P]/dt = 0$ (C1a)

Note again, as discussed above, that the $\dot{e}$ is the energy exchange by the bird, $q^{\cdot}$ is the rate, of heat exchanged (or produced), and $(-w)^{\cdot}$ is the work done by the bird. The term $d[(V\Delta P)]/dt$ captures the change in energy not captured in the enthalpy balance of an open system. The *enthalpy balance* is considered below without this term in the overall energy balance. The change in kinetic energy is zero between the exit plane and inlet plate at a steady-state (the control volume is fixed in space). The subscript (LB) signifies a lone bird. At steady-state:

$[e^{\cdot}]_{LB} = q^{\cdot} - w^{\cdot} = m^{\cdot}C_p.(T_1-T_0)$ (C1b)

Here V (volume)=$L^3$. The control volume has fixed dimensions. Assume that $d[(V\Delta P)]/dt$ is small (we also approximate that all the mass flow of air is in the horizontal flight direction although there will be some in the z-direction). Such terms may also cancel out in the comparisons below for determining the optimal formation and so are ignored for this model. Air is an ideal gas. An entropy balance at steady-state yields, the rate of entropy generation:

$[\phi^{\cdot}]_{LB} = m^{\cdot}C_p\,[\ln(T_1/T_0)] - m^{\cdot}R.[\ln(P_1/P_0)]$ (C1c)

Here R is the gas constant equal to ($C_p$- $C_v$) for an ideal gas. The second term is at least about 20% smaller than the first term for similar logarithmic groupings. The entropy generation $[\phi^{\cdot}]_{H3}$ is equal to the entropy leaving the volume minus that entering it. For the control volume type of calculation (from reference 4), the rate of entropy generation can be approximated as:

$[\phi^{\cdot}]_{LB} = m^{\cdot}C_p\,[T_1-T_0]/T_0 - m^{\cdot}R.\ln(P_1/P_0)]$ (C1d)

**Case 2: H-Formation, 3 Birds (Figure 1 (b)).**

Energy Balance at steady state:

$3[e^{\cdot}]_{H3} = 3[\dot{q}] - 3[w^{\cdot}] = 3[m^{\cdot}C_p.(T_1-T_0)]$ (C2a)

Assume that the control volume is at a fixed entropy i.e., at steady state.

$[\phi^{\cdot}]_{H3} = 3\,[\phi^{\cdot}] = 3[m^{\cdot}C_p\,[\ln(T_1/T_0) - m^{\cdot}R.\ln(P_1/P_0)]$ (C2b)

**Case 3: I-Formation, 3 Birds (Figure 1 (c)).**

Energy Balance:

$3[e^{\cdot}] = 3[q^{\cdot} - w^{\cdot}]_{I3} = m^{\cdot}.C_p.(T_3-T_0)$ (C3a)

Assume that the control volume is at a fixed entropy i.e., at a steady-state. The entropy generation $(\phi^{\cdot})_{I3}$ is equal to the entropy leaving the volume. Thus:

$[\phi^{\cdot}]_{I3} = m^{\cdot}C_p\,[\ln(T_3/T_0)] - m^{\cdot}R.[\ln(P_3/P_0)]$ (C3b)

Note that $T_3$ will always be greater or equal to $T_2$.

**Case 4: Separated V- Formation, 3 Birds with the partitioned temperature at the exit plane. (Figure 1(d)).**

*Energy Balance:* x/L is the fraction overlap as shown in Figures 1 and 3. Assume $E^{\cdot} = 3e^{\cdot}$. The energy balance becomes,

$3[e^{\cdot}]_{SV3} = m^{\cdot}C_p.\,[2(1-x/L)\,(T_1-T_0) + (1-2x/L)\,(T_1-T_0) + 2(x/L)\,(T_2-T_0)]$ (C4a)

$3[e^{\cdot}]_{SV3} = 3[q^{\cdot}]_{SV3} - 3[w^{\cdot}]_{SV3} = m^{\cdot}C_p\,[(3-4x/L)\,(T_1-T_0) + 2(x/L)\,(T_2-T_0)]$ (C4b)

*Because the differentiation of the energy with distance is the force*, and there is no force in the lateral (x) direction, one can write (for a fixed $T_1$ and $T_0$):

$d[e^{\cdot}]_{SV3}/d(x/L) = -4(T_1-T_0) + 2(T_2-T_0) + 2(x/L)\,(d(T_2-T_0)/d(x/L) = 0$ (C4c)

It should be noted that there are no measurements available that adequately can be used to validate most of the common avian flight models. Regardless, there are temperature bounds that can be inferred from careful wing tunnel measurements made on individual birds [34-42]. The following sets of temperatures are used in the calculations to illustrate the temperatures and efficiencies. The sets are at best guess except for the $T_0$ temperature which appears to be a fair approximation for the temperature at the altitudes of geese flights. The two sets represent low and high-temperature excursions (i) $T_0$=275K, $T_1$= 276K, and (ii) $T_0$=275K, $T_1$= 285K respectively. Note that the two cases may represent higher than the actual temperatures which will depend on the converted chemical energy, however, the trends and conclusions are not affected.

For the case of no-interaction i.e., at x/L=0, $T_2$ must approach $T_1$. The plot of $T_2$ vs x/L can be made if $T_0$ and $T_1$ are known (measured). If all terms of the entropy balance are known and if the entropy generation term is at an extremum, then $T_2$ can be inferred from Equations (C1b and C1c). Regardless, an approximate solution for equation (C4c) takes the following form [$T_2=2T_1-T_0-K/(x/L)$] which for the set (i) approximately can be written in the following form:

$T_2=-K/(x/L) + 277$ (C4d)

Figures 4 (a) and (b) are plots of $T_2$ for the two sets. Here K is a constant. Assuming $T_2=T_1$ at x=0 does not yield a unique positive value of K, however, assuming $T_2=T_1$ when x is very small can yield a solution for x/L>0 where K is small compared to $T_0$. The calculated small value for K will slightly change the absolute value of $T_2$ (particularly when $T_1$ and $T_0$ are close) but not the trend or shape of the $T_2$ as a function of x/L. For obtaining a solution we have assumed $T_2=T_1$ when K is small approximated as K=0.0001. K is the overlap constant.

***The choice of K at an arbitrarily small value of x/L is an approximation that impacts the slope of the curves in Figures 5 (a) to (c), which allows for the matching with the experimental observations (this is discussed in Section 4). There is some evidence to suggest that wing morphologies can be adjusted to avoid this kind of potential theoretical divergence. [10,36].*** A physical rationale for the approximation is not offered in the article except to note that wingtips are often not smooth for several avian species, a condition that could perhaps establish the smallest non-zero x/L where the solution is possible. It is fully recognized that the solution (represented by equation (C4d) could be divergent as x/L approaches 0).

For simplifying the entropy rate comparisons, we approximate that $P_1$~$P_2$~$P_3$ are equal and all very close to the $P_0$ i.e., at the flight altitude. Assume that the control volume has a fixed entropy i.e., at a steady-state, the entropy generation $[\phi^\cdot]_{SV3}$ is equal to the entropy leaving the volume.

$[\phi^\cdot]_{SV3} = [(3-4(x/L) (m^\cdot.C_p.\ln(T_1/T_0) +2(x/L) (m^\cdot C_p.\ln(T_2/T_0)]$ (C4e)

To compare the entropy generated per unit-bird as a function of the power expended, we define $\beta(K^{-1})$ as the entropy generation rate per unit rate energy expended. For the three bird SV formation,

$\beta=[(3-4(x/L)) (m^\cdot.C_p.\ln(T_1/T_0)-m^\cdot R.(P_1/P_0)] + 2(x/L) (m^\cdot C_p.\ln(T_2/T_0)-m^\cdot.R.(P_2/P_0)]/[(3L-4(x/L))m^\cdot.C_p.(T_1-T_0) + 2(x/L).m^\cdot C_p.(T_2-T_0)].$ (C4f)

__Case 5. ST- Staggered formation, 3 Bird. Figure 1(e).__

$[E^\cdot]_{ST3}= [-2(1-x/L)\, m^\cdot.\, C_p.(T_1-T_0) – (1-2x/L)m^\cdot.\, C_p.(T_1-T_0)-2(x/L)\, m^\cdot C_p.(T_2-T_0)] +3e^\cdot]$ (C5a)

The entropy generation of the three birds in the staggered formation is:

$[\phi^\cdot]_{ST3} =[2(1-x/L) (m^\cdot.\, C_p.(T_1-T_0)/T_0 -m^\cdot R.\ln(P_1/P_0)] + [(l-2x)\, m^\cdot.\, C_p.(T_1-T_0)/T_0 - m^\cdot R\ln(P_1/P_0)] + [2x/L\, m^\cdot C_p.(T_{2s}-T_0)/T_0-m^\cdot R.\ln(P_2/P_0)]$ (C5b)

Here $T_{2S}$ is $T_2$ for the staggered.

At x/L=0.5

$3e^\cdot= m^\cdot.\, C_p.(T_1-T_0) +m^\cdot C_p.(T_2-T_0)]$ (C5d)

$[\phi^\cdot]_{ST3} = [m^\cdot.\, C_p.(T_1-T_0)/T_0 -m^\cdot R(P_1/P_0)] + [m^\cdot C_p.(T_2-T_0)/T_0) - m^\cdot R\ln(P_{2s}/P_0)]$ (C5e)

$P_{2s}$ is $P_2$ for staggered.

__Case 6. SV- Formation, 5 Birds. Figure 3(a) .__

*Energy Balance:*

$5[e^\cdot]_{SV5}= [2(L-x)\, m^\cdot.C_p.(T_1-T_0) + 3(L-2x)m^\cdot C_p.(T_1-T_0) + 4x.m^\cdot C_p.(T_2-T_0)]/L)$ (C6a)

At x/L=0.5

$[e^\cdot]_{SV5}= (1/5)\, m^\cdot C_p\, [ (T_1-T_0) +2 (T_2-T_0)]$ (C6b)

*Entropy Generation:*

$[\phi^{\cdot}]_{SV5} = [(5L-8x)\, m^{\cdot}.C_p.\ln(T_1/T_0) + 4x.m^{\cdot}C_p.\ln(T_2/T_0)]/L$ (C6c)

At x/L=0.5

$[\phi^{\cdot}]_{SV5} = [m^{\cdot}.\, C_p.(T_1/T_0-1) - m^{\cdot}R.\ln(P_1/P_0)] + 2[m^{\cdot}C_p.(T_2/T_0-1) - m^{\cdot}R\ln(P_2/P_0)]$ (C6d)

$[\phi^{\cdot}]_{SV3} = [(m^{\cdot}.C_p.(T_1/T_0-1) - m^{\cdot}R.\ln(P_1/P_0) + m^{\cdot}C_p.(T_2/T_0-1) - m^{\cdot}R\ln(P_2/P_0)]$ (C7d)

**Case 7. ST- (Staggered) Formation, 5 Birds. Figure 3(c).**

*The staggered subscript used in Case 5 above is dropped.*

*Energy Balance:*

$5[e^{\cdot}]_{ST5} = [2(L-x)m^{\cdot}.C_p.(T_1-T_0) + 3(L-2x)\, m^{\cdot}.C_p.(T_1-T_0) + 4x\, m^{\cdot}C_p.(T_2-T_0)]/L]$ (C7a)

*Entropy Generation:*

$[\phi^{\cdot}]_{ST5} = [(5L-8x)\,(m^{\cdot}.C_p.\ln(T_1/T_0) - m^{\cdot}R.\ln(P_1/P_0)) + 4x((m)^{\cdot}C_p.\ln(T_2/T_0) - m^{\cdot}R.\ln(P_2/P_0)]/L$ (C7b)

At x/L=0.5,

$[\phi^{\cdot}]_{ST5} = [m^{\cdot}.C_p.(T_1-T_0)/T_1 - \dot{m}.R.\ln(P_1/P_0) + 2\dot{m}.C_p.(T_2-T_0)/T_2] - 2\dot{m}.R.\ln(P_2/P_0)]$ (C7c)

This is identical to the SV formation for the same number of birds in the flock.

**Case 8. SV Formation, 7 Birds.**

*Energy Balance:*

$7[e^{\cdot}]_{SV7} = [2(1-(x/L))m^{\cdot}C_p.(T_1-T_0) + 5(1-2(x/L))\,\dot{m}.C_p(T_1-T_0) + 6(x/L)m^{\cdot}.C_p(T_2-T_0)]$ (C8a)

$7[e^{\cdot}]_{SV7} = [(7-12x/L)m^{\cdot}.C_p.(T_1-T_0) + 6(x/L)m^{\cdot}.C_p.(T_2-T_0)]$ (C8b)

*Entropy Generation:*

$[\phi^{\cdot}]_{SV7} = [(7-12x/L)\,(m^{\cdot}.C_p.\ln(T_1/T_0) - m^{\cdot}.R.\ln(P_1/P_0)] + 6x/L[m^{\cdot}C_p.\ln(T_2/T_0) - m^{\cdot}R.\ln(P_2/P_0)]$ (C8c)

At x/L=0.5

$7[e^{\cdot}]_{SV7} = [m^{\cdot}.\, C_p.(T_1-T_0) + 3m^{\cdot}C_p.(T_2-T_0)]$ (C8d)

$d[e^{\cdot}]_{SV7}/d(x/L) = -12(T_1-T_0) + 6(T_2-T_0) + 6(x/L)\,(d(T_2-T_0)/d(x/L) = 0$ Same as 3 bird force

For the entropy generation rate for a 7 birds SV flock at x/L=0.5,

$[\phi^{\cdot}]_{SV7} = [\,(m^{\cdot}.C_p.\ln(T_1/T_0) - m^{\cdot}.R.\ln(P_1/P_0)] + 3[m^{\cdot}C_p.\ln(T_2/T_0) - m^{\cdot}R.\ln(P_2/P_0)]$ (C8e)

**Case 9. SVT Tight Formation 7 birds, Figure 3(d).**

*Note that $T_3$ is always higher or equal to $T_2$.*

*Energy Balance:*

$7[e^{\cdot}]_{SVT7} = [\,(3-4x/L)\, m^{\cdot}.C_p.(T_1-T_0) + 4(x/L)\,\dot{m}.C_p.(T_3-T_0) + 2(1-2x/L)m^{\cdot}.C_p.(T_2-T_0)]$ (C9a)

At x/L=0.5

$7[e^{\cdot}]_{SVT7} = [m^{\cdot}.C_p.(T_1-T_0) + 2\,\dot{m}.C_p.(T_3-T_0)]$ (C9b)

*Entropy Generation:*

$[\phi^{\cdot}]_{SVT7} = [m^{\cdot}.C_p.\ln(T_1/T_0) - m^{\cdot}R\ln(P_1/P_0) + 2m^{\cdot}C_p\ln(T_3/T_0) - 2m^{\cdot}R.\ln(P_3/P_0)]$ (C9c)

The energy balance and entropy generation can be extended to a larger flock of birds.

# 4. Discussions.

Although experimental observations and measurements of avian flight are not adequate to test any flight-model rigorously, it must be mentioned that some of the measurements reported to date are quite remarkable given the difficulty of obtaining them. The main experimental observations from the published literature [1,2,3,9,10] include:

(i) The "V" formation (SV) and staggered (ST) formations are the most observed formations.
(ii) Formation flying is advantageous for energy efficiency.
(iii) The energy expenditure per bird is reduced with increasing flock size.

Based on the model presented in Section 3., Figures 4(a) and (b) are the plots of $T_2$ vs x/L with the two sets of temperatures considered and K=0.0001. The $T_2$ temperature increases with x/L as it is related to the fraction of the overlap of the trailing bird with the leading bird, with the constraint of no lateral force for the formations considered. We note that trends are not altered by the specific choice of temperatures of $T_0$ or by the quantity of work produced.

The energy expended [$\dot{e}$] as a function of x/L can be calculated and plotted with the $T_2$ temperatures obtained from the model. The main methods to improve the efficiency of any operation in a typical thermodynamic closed-cycle loop are to increase the *mean* temperature of energy addition and reduce the *mean* temperature of energy rejection which is what happens in the model. From the formation examples (Case 2 – Case 9), it is seen that the highest $T_2$ is always obtained at x/L=0.5 (half coverage in the SV and ST configurations). The lowest $T_2$ temperature is $T_1$, i.e., when x/L= 0 (no coverage with the assumption that any divergence at x/L=0 is treated as discussed above). The efficiency ratio as a function of x/l, is plotted in Figures 5(a), (b), and (c) for the two sets of temperatures for a 3-Bird and 7-bird SV formation. The energy usage in a formation (per bird), decreases when compared to the lone bird. Regardless, one should note that the Carnot efficiency for heat to work (Equation 1), cannot be exceeded. This efficiency could be quite low, given the small difference between high and low temperatures.

***The model proposed in this article agrees with all three main experimental observations. However, the assumption of a constant in equation (C4d) should be noted for the acceptance of the efficacy of this model.***

It is not clear how the flock of birds selects the altitude they fly at. The density of air and drag resistance falls with altitude as does the temperature (excepting cases where there is an atmospheric inversion). Presumably, the altitude selection offers the lowest drag, lowest headwinds, and the ability to consistently perform the wing work required for flight while avoiding predators and obstacles.

For birds, there may also be a need to maintain a certain core body temperature for comfort, particularly during long flights. From Figures 4(a) and 4(b), we note that the calculated $T_2$ temperatures at x/L =0.5 approach 277K and 295K respectively. If we assume that the drag reduces with altitude as does the outside temperature, then the fact the temperature can increase for the maximum coverage (e.g., x/L=0.5), as noted in Figures 4(a) and 4(b), is perhaps a reason to assume that the birds fly at the highest altitude they can, which give them comfort while reducing drag. The model thus allows for flight at high altitudes because the thermal conditions of birds in an optimal formation are favorable for conserving energy at a higher temperature. The plots of the entropy generation rate and the entropy generation rate per unit of energy are shown in Figures 6(a) and (b).

Birds have high basal metabolic rates and so use and transfer chemical-free energy at high rates [39,40]. Oxygen intake is known to be a steep function of the outside temperature i.e., drops significantly with temperature [35.36]. By choosing a low $T_0$ temperature (e.g., with a higher altitude flight) the metabolic rate and heat production can be simultaneously decreased, thereby possibly retaining the "fuel" for a more efficient work expenditure.

Heat is produced during the flight. In the model presented in this article, the trailing birds use some of the heat from the leading bird(s) and produce less heat or convert thermal energy to work, thus garnering efficiency in energy deployment. The energy expended per bird per unit of time is lower with increasing flock size as can be noted by comparing Equation (C6a) or (C8a) with Equation (C4a). Figures 5(a), (b), and (c), indicate that increasing the flock size or the $T_1$ temperature, lowers the amount of energy in use. We note that the energy production rate per bird in the SV Equation (C4a) is lower when compared to other formations like the H formation Equation (C2a)) and possibly the I formation (Equation (C3a)). Note that $[\dot{e}]_{I3} < [\dot{e}]_{H3}$. For the H formation, there is no benefit from energy savings over the individual flight per bird for the same velocity of flight. Regardless of the energy savings in the, I formation, namely in Case 3, the entropy generation rate is lower than that in Case 2. As $T_2$ is always greater or equal to $T_1$, a comparison of energy used or the same work (i.e., for the same velocity) yields $[\dot{e}]_{SV} < [\dot{e}]_{H}$. Compared to the I formation the SV3 formation will always indicate a lower energy expenditure per bird as the flock size increases (in the I-formation the trailing bird temperature always increases unlike in the SV formation case). However, every trailing bird in the I-formation is progressively warmer, e.g., the third bird will be warmer than the second bird and so on which is not sustainable for the comfort of the bird (but can explain the rotation of positions). The entropy generation of the H formation is easily argued to be higher than the I formation unless ($T_3$-$T_0$) is equal to 3($T_1$-$T_0$) which would effectively mean that the birds do not interact. The maximum x/L cannot exceed 0.5 for a uniform pattern to develop. It has already been discussed that the values of the $T_0$ and $T_1$ temperatures, when fixed, change the actual temperature of $T_2$ and not the trend of $T_2$ vs x/L for a fixed value of K. The power expended per bird can be plotted as a function of the flock size by comparing it to the three-bird case. This plot is shown in Figure 6(a) for the SV formation. From Figure 6(b), we note that in the SV formation, the total power expended per bird falls with an increasing number of birds in the flock (when compared at equal flock velocities).

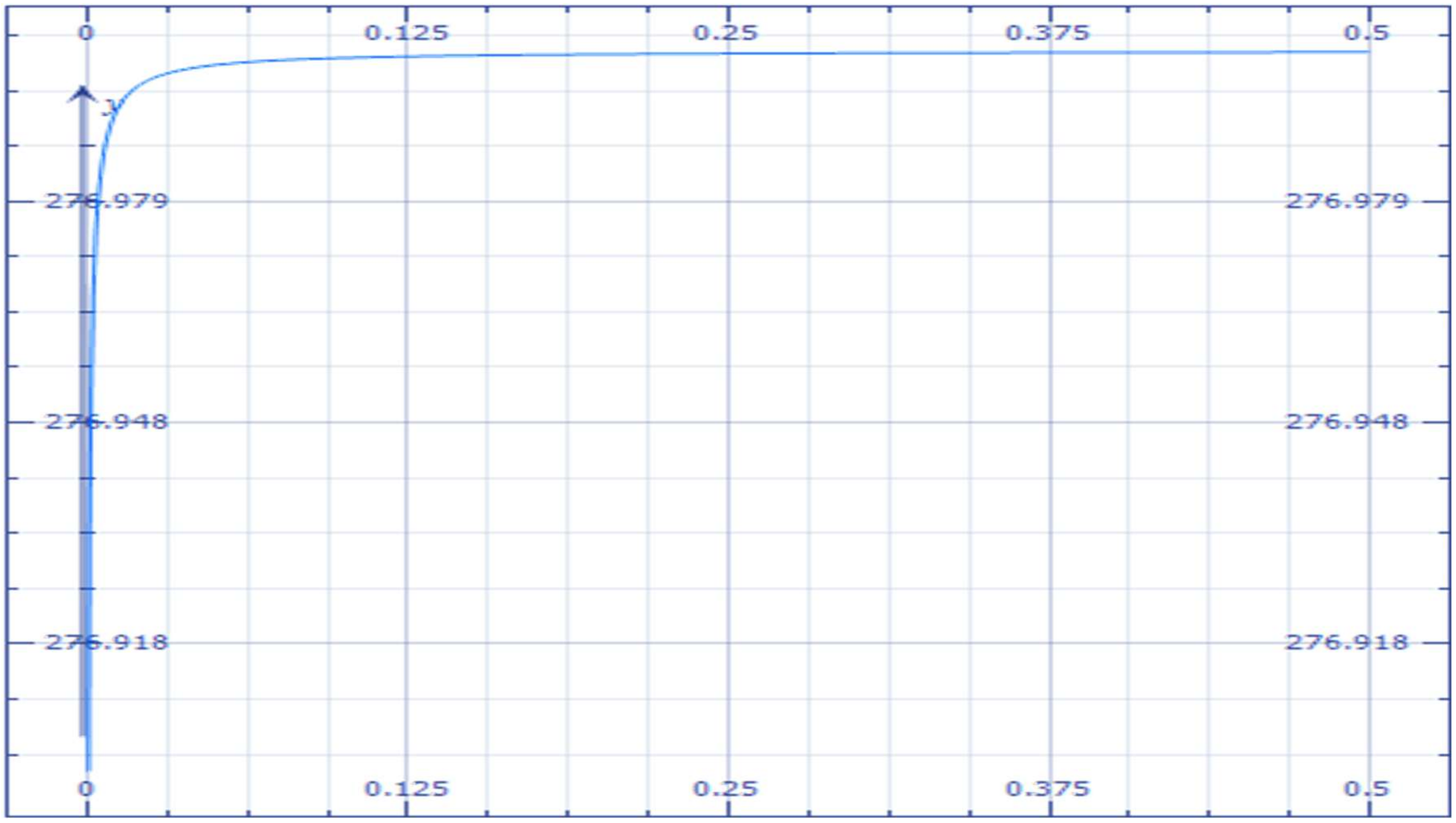

**Figure 4(a).** The plot of $T_2$ vs x/L when $T_0$ and $T_1$ are fixed for Case 4 (3 Bird SV). The temperature set used is set (i) i.e., $T_0$=275K, $T_1$= 276K, and $T_2$=$T_1$ at x/L~0. K=0.0001.

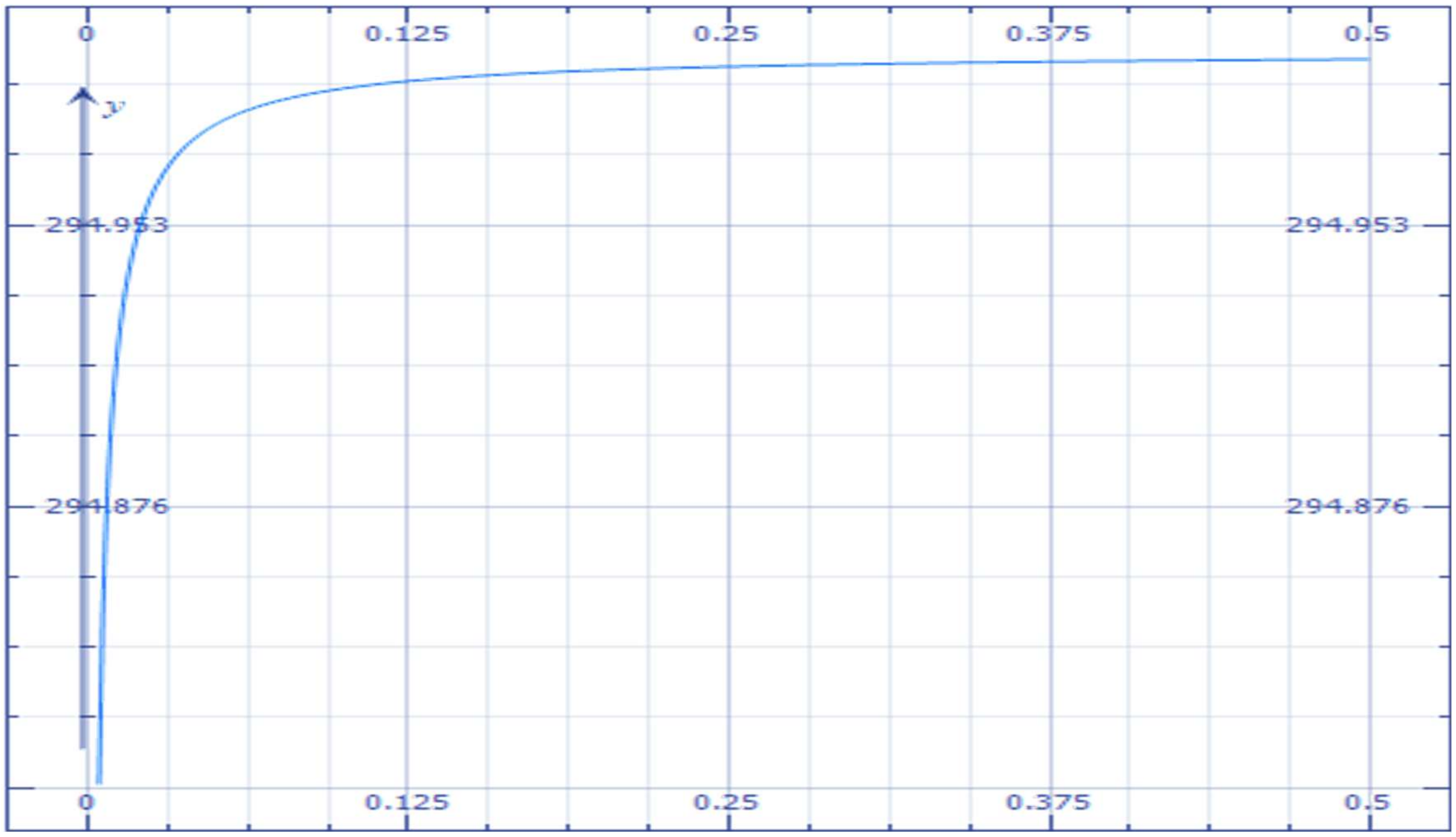

**Figure 4(b).** The plot of $T_2$ vs x/L for Case 4. The temperature set employed is set (ii) i.e., $T_0$=275K, $T_1$= 285K, and $T_2$=$T_1$ at x/L~0. K=0.0001.

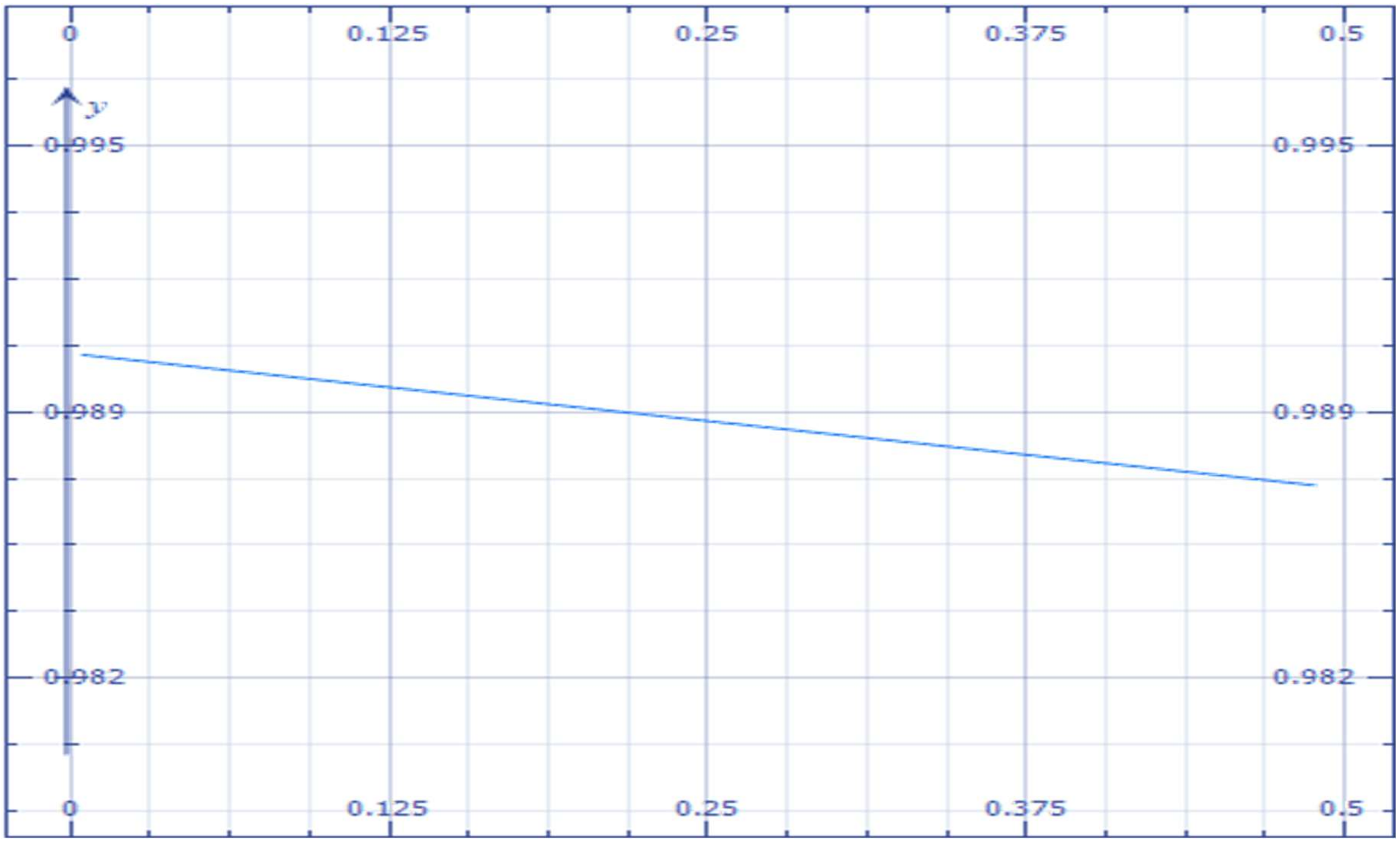

**Figure 5(a).** A plot of $[e^{\cdot}]_{SV3}$ /$[e^{\cdot}]_{LB}$ per bird as a function of x/L. The temperature set employed is $T_0$=275K, $T_1$= 276K, and $T_2$=276.99K at x/L=0.5. K=0.0001.

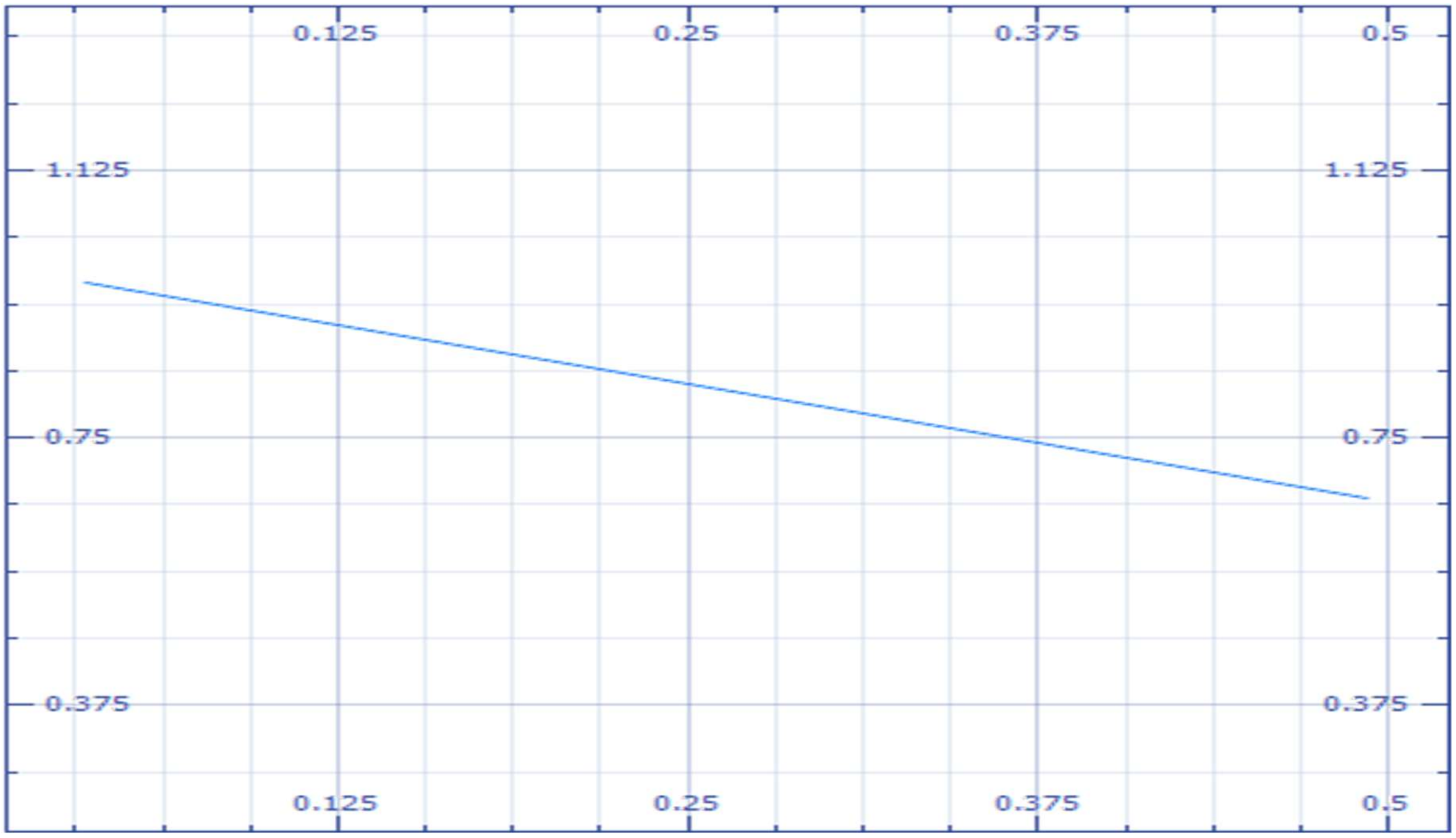

**Figure 5(b).** A plot of $[e^{\cdot}]_{SV3}$/ $[e^{\cdot}]_{LB}$ per bird as a function of x/L. The temperature set employed is $T_0$=275K, $T_1$= 285K. K=0.0001.

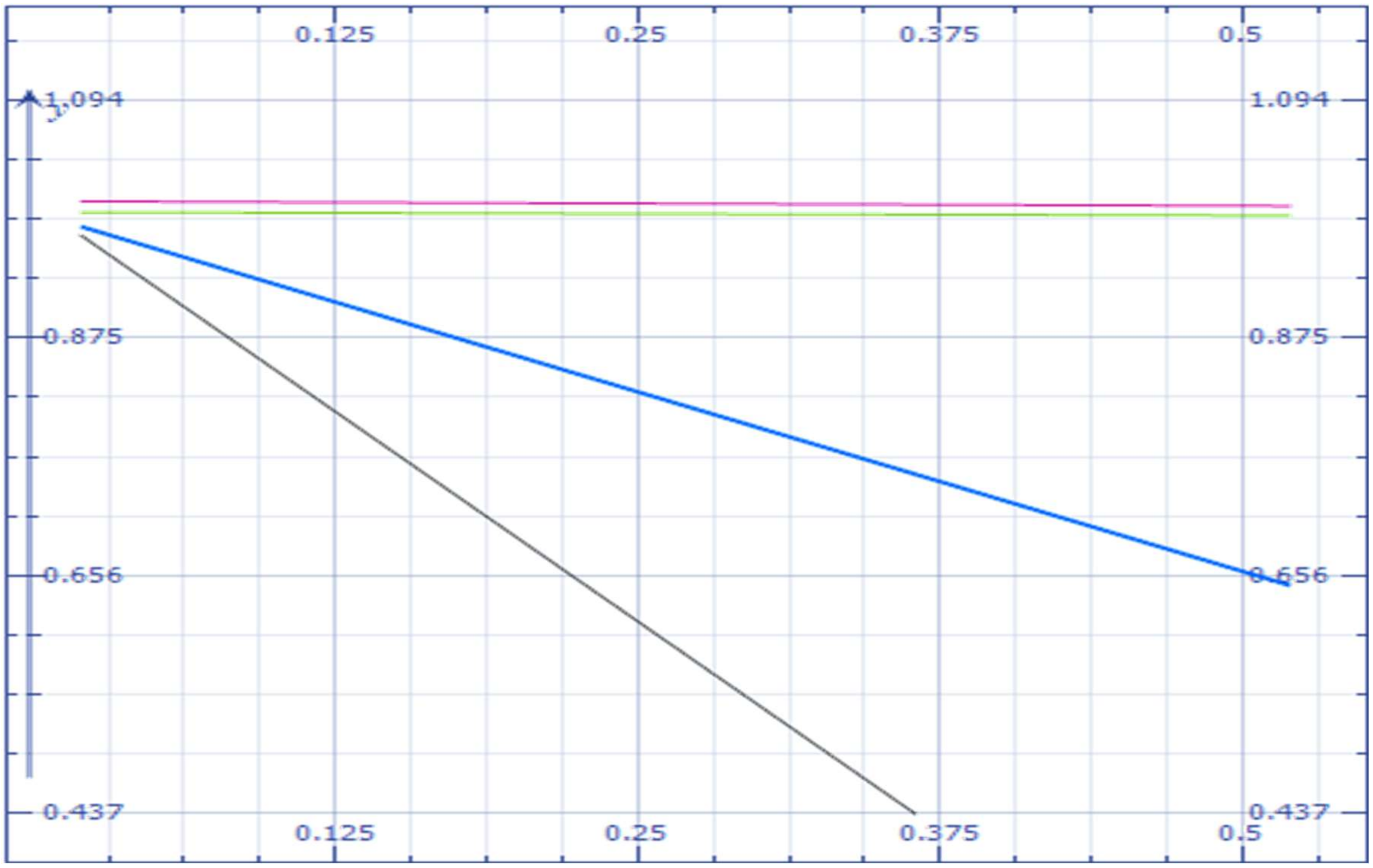

**Figure 5(c).** A plot of the 3-Bird $[e^{\cdot}]_{SV3}/[e^{\cdot}]_{LB}$ and 7 bird $[e^{\cdot}]_{SV7}/[e^{\cdot}]_{LB.}$ as a function of x/L. The top two lines red (3-Bird) and green (3-Bird) are for Set (i) and the bottom two lines (blue (7-Birds) and black (7-birds) are for Set (ii). K=0.0001.

The "V" formation (SV formation) shows the maximum rate of entropy generation along with the staggered ST formation. Equations (C4e) and (C4f) are used as examples of equations to understand the trend of the entropy generation rate and entropy generation rate per unit power, as a function of x/L, for the SV formation. The plots of the solution for the rate are shown in Figure 7 for the 3 Bird SV formation for both sets of temperatures (i) and (ii). Regardless of the sets used, the rate is maximum at x/L=0.5. To compare the entropy generation rate between the various possible formation (patterns), it must also be shown that the entropy generation rate is at a maximum for all x/L. This is easily inferred because the second differential of the entropy generation with x/L for Cases (2-9) is always positive, thereby establishing the maximum entropy generation rate for a particular formation of flight. The entropy generation w.r.t. (x/L) for a fixed $T_0$ and $T_1$ is always positive for all flock sizes, (equation (C8c) for example) if $T_1>T_0$. Note that the best free-energy conversion efficiency for a bird will approach $(1-T_0/T_H)$, the Carnot-efficiency limit. Here $T_H$ is the maximum temperature in the bird core which is a stable temperature for every bird species [10]. Consequently, a comparison of the rates between the various patterns as made in Section 3, for establishing the maximum entropy generation rate between the various formations is a valid method for comparing the entropy generation rate between various patterns.

Note that $T_3$ is always greater or equal to $T_2$. The various formations (Case 2 to 9) can also be compared for entropy generation by comparing the rates at x/L=0.5. For example, Equation (C8c) predicts a higher number compared to Equation (C9c) if $T_3-T_0=3(T_1-T_0)$. The SV formation indicates the highest entropy generation rate.

Based on the results and the experimental observations it appears that the MEPR postulate could thus be valid for predicting avian flight formation. This is a result that validates the self-organization principles discussed in reference 4. Regardless, it should be noted that not all possible formations have been compared. Smaller birds like sparrows do not fly in an in-plane “V” formation, nor do they fly such long migratory distances without stopping and feeding. However, sparrows do fly in a formation that is multilayered and tapered vertically. On the other hand, starlings establish very clear patterns of flight even over small distances flown. A similar model in three dimensions may perhaps be applicable for the smaller birds. This is left to a future study.

As shown in Figure 8, for the 3-bird SV formation, the x/L~0.5 position displays the highest $\beta$**, (**the ratio of entropy generation per unit power). For the MEPR selection, we are interested in comparing the value at any x/L *between* various formations. Although not calculated explicitly in this article because the entropy generations can be compared directly, as shown above, likely, a comparison of $\beta$ for all the formations will also be a validation of MEPR.

Experimentally it has been recorded that the energy expended per bird per unit time is always lowered both by the “V” formation *and* with increasing flock size [1,2,3,9,10,24]. Both the SV and ST formations appear equally preferable in the model unless the lead bird in the ST formation produces less energy, in which case the SV will be preferred over the ST from an overall energy balance. If the staggard has a temperature of $T_{2s}$ which is different than $T_2$ for the SV. From Equation (C5e) we note that, if $(T_{2s}-T_0) < 2(T_1-T_0)$ and the pressure terms cancel, the entropy generation is lower than Case 1 for this staggered formation. When comparing Equations (C5a) and (C4a) note that Case 5 has the same entropy generation as Case 4 for the same x/L when $T_{2s}= T_2$. However, there is a difference, because the first bird can only part with half the heat being transferred from the leading bird to a trailing bird, and half is wasted. However, as the flock grows, this difference will diminish this making the staggard ST and SV predict very similar entropy generation ate and energy savings.

A difference between the eddy-lift models [12,3,9] and the thermal model of this article is that the birds must overlap in this model to realize energy benefits whereas perhaps not required in the eddy lift models. However, if the relative wind is angled then the overlap condition is not as clear. In the present model, the entropy generation is only from the thermal gradients. Some entropy generation is possible from local pressure gradients which have not been factored into the present model. Should these pressure gradients also be the highest entropy generating conditions for the V formation then the MEPR postulate will be comprehensively established as a possible behavioral principle for anticipating the self-organization behavior of live systems.

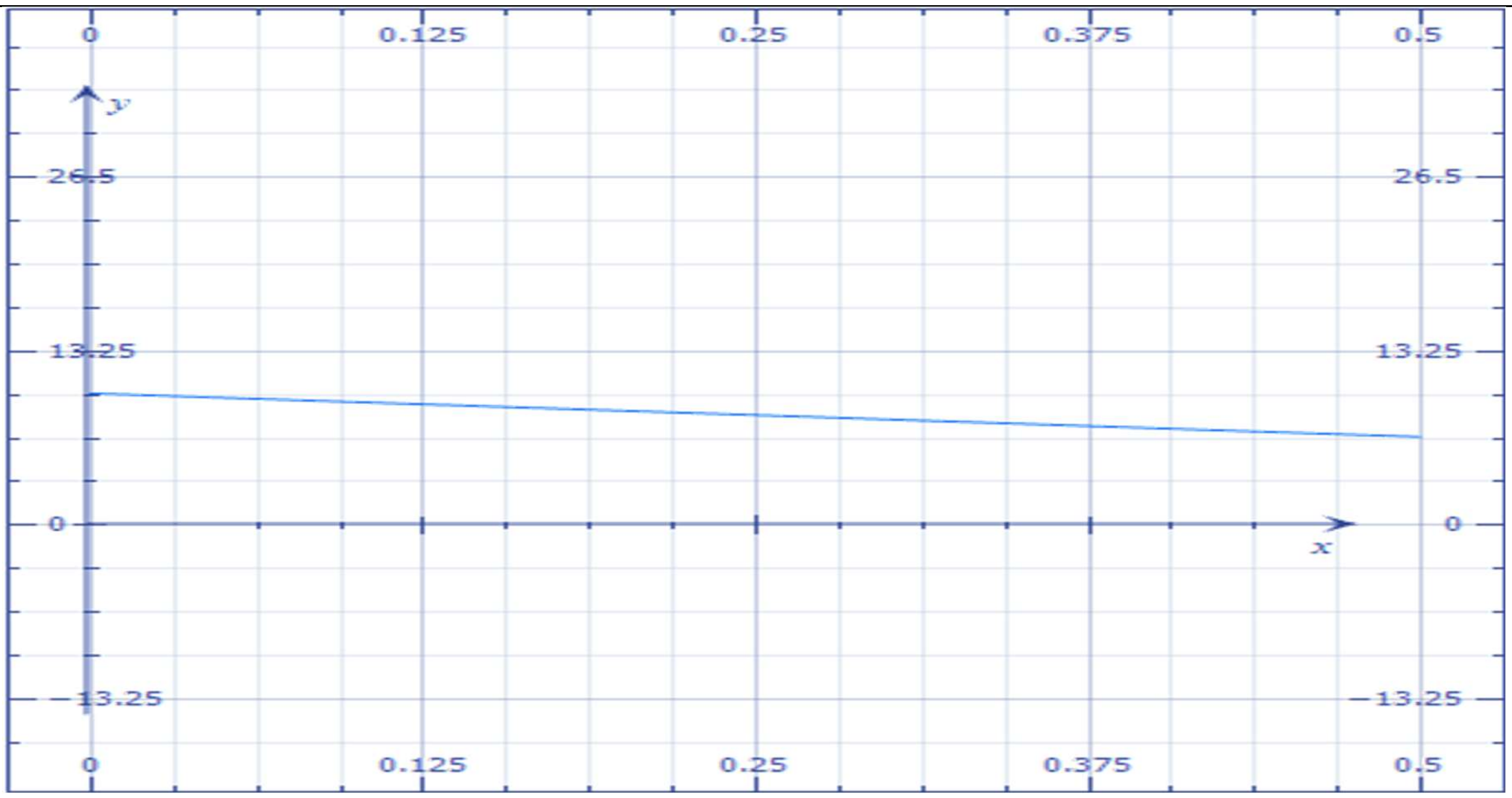

**Figure 6(a).** The ratio of Energy expended per bird in a 7 bird flock and 3 bird flock as a function of x/L for the lower T1 (276C). T0-275C. Temperature Set (i). K=0.0001.

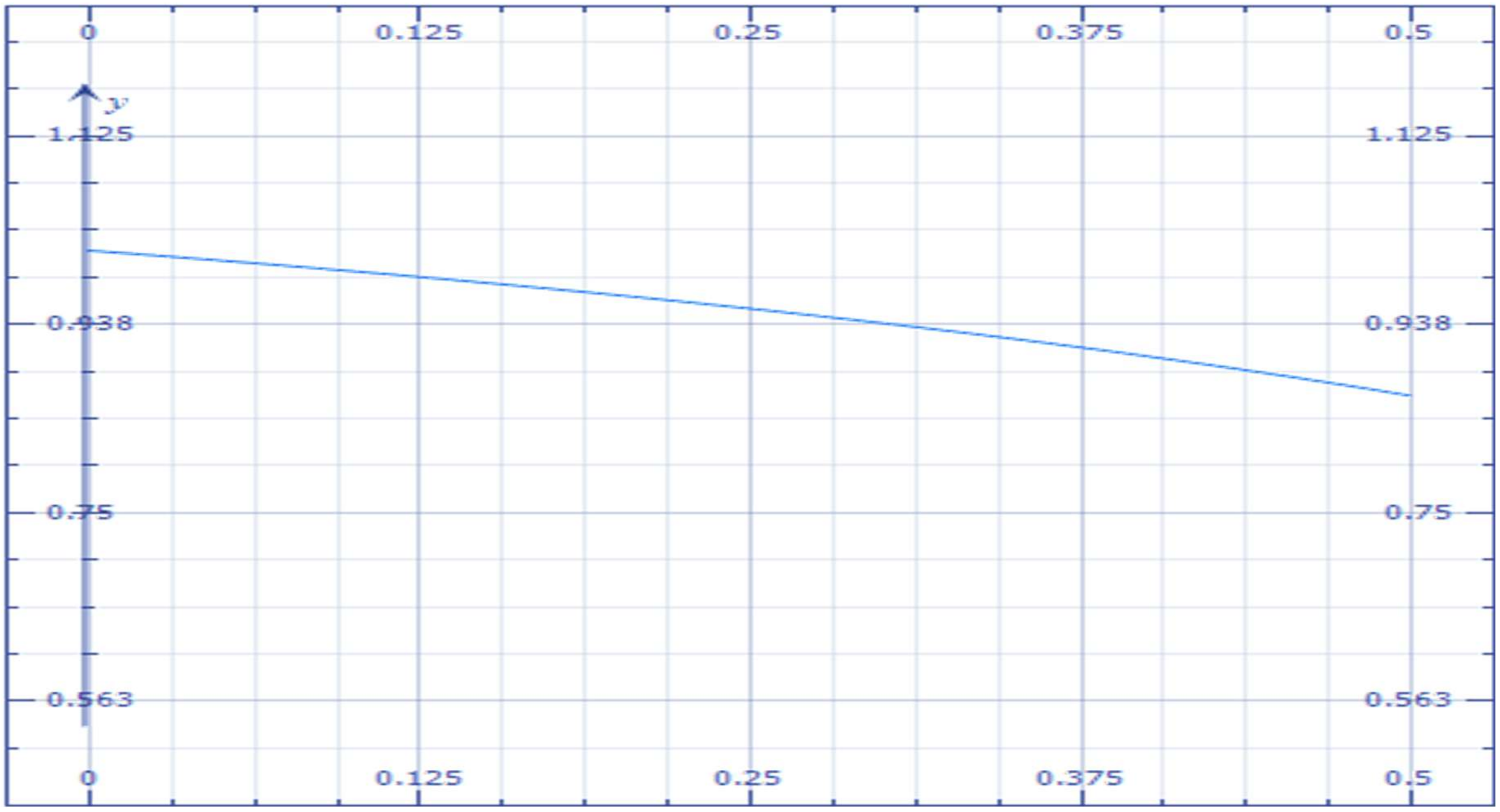

**Figure 6 (b).** The ratio of Energy expended per bird in a 7-bird flock and 3-bird flock as a function of x/L for fixed $T_0$ and $T_1$ corresponding to Set (ii). K=0.0001.

Figure 7 is an integrated plot for entropy generation rate vs x/L. for 3-Birds for both sets (i) and (ii). Figure 8 is a plot β ($K^{-1}$) namely, the entropy generation rate **divided** by the rate of energy expenditure per bird as a function of x/L for a 3 Bird flock. $T_0$=275K, $T_1$=285K, and $T_2$~T1 at x/L~0.

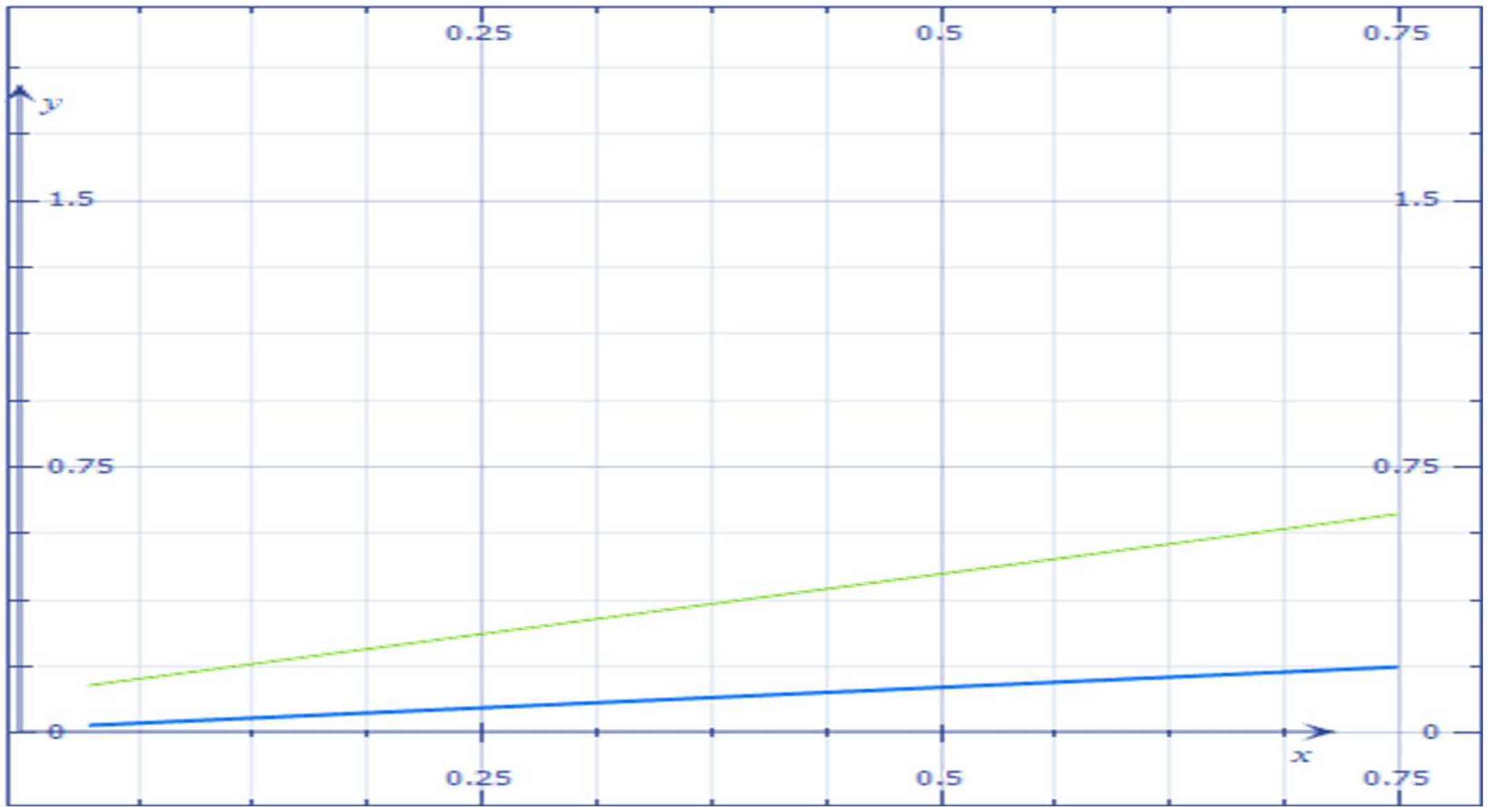

**Figure 7.** Integrated 3 Birds Set (i) and (ii) for Entropy Generation Rate vs x/L. K=0.0001.

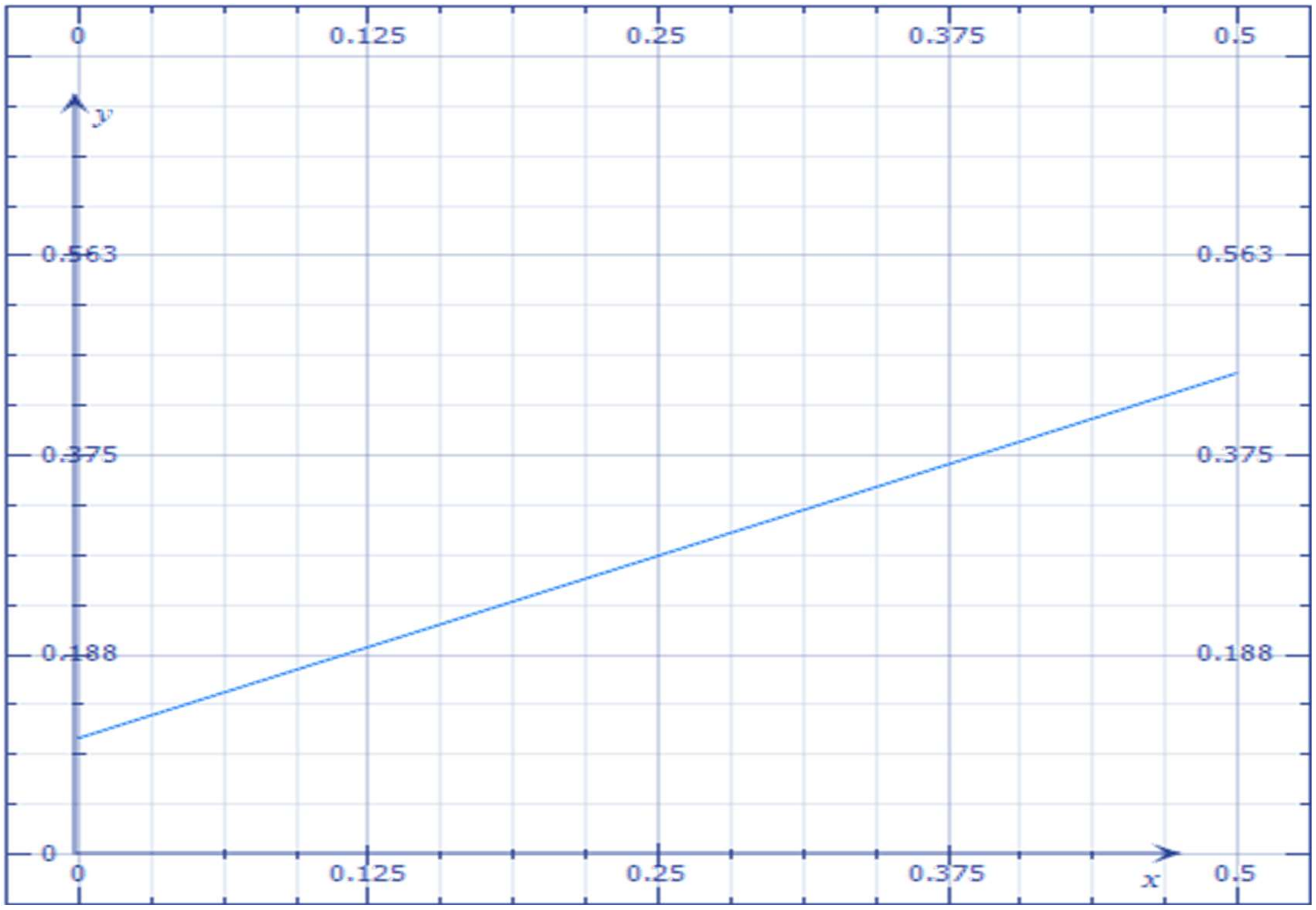

**Figure 8.** A plot β ($K^{-1}$) (Entropy generation rate **divided** by the rate of energy expenditure per bird) as a function of x/L for a 3 Bird flock. $T_0$=275K, $T_1$=285K, and $T_2$~T1 at x/L~0. K=0.0001.

*Large birds are possibly extremely sensitive to temperature change* [10,34]. Formation-flying requires communication and assessment of position both visually and by the sensors of force and temperature. When birds exchange places (in Reference 1 this is expected to happen when the lead bird has expended 50% of the energy) it must do so in a fashion that minimizes the possibility of a pattern break-down.

Treating the information exchange as a discrete decay-prone process compared to the overall system, the entropy production when at an extremum is a logical way to assume low disruption. It is likely that pattern disruptions, are quickly sensed by the birds' thermoreceptors, and can thus be tempered if unbearable. It is unlikely that the bird recognizes energy efficiency from a pattern, but the bird can certainly sense temperature very accurately [34-42] and fly at a comfortable temperature a feature discussed above. Investigations of thermo-reception and thermoregulation in birds indicate that thermo-sensors exist not only in the skin on the body but also in the skin around the face, the thoracic brood patch (the critical area used for egg incubation), and the beak region [34-42]. Most birds are homeothermic, normally maintaining their body temperature within a range of less than a few degrees Kelvin by active metabolic means. For mammals, the high degree of development for sensing temperature provides them with the capacity to use temperature or thermal information, not only as a signal of the condition of the body but also as a sensory input, useful for recognizing objects and exploring the environment. Should the same hold for birds, then there is a fair expectation that breaking the formation could be tied to an uncomfortable temperature increase. Although by no means definitively known, it appears that birds do not learn to fly in the optimal formation because of a genetic information transfer, but rather learn about flying in a formation by experience and possibly by communicating with other birds [1,2,6,10,21,24, 34,35,36,37,38,39,40,41,42] i.e., they train themselves to do so. Temperature sensing may thus play a significant role in the choice of a particular pattern.

Large birds communicate and dissipate some heat through their moth regions sometimes with loud honks in short sound-packets during flight (somewhat akin to panting in mammals). There is a need for more experimental data on the energy exchange process and temperature variations that are associated with such information exchange which may help the understanding of why the birds cooperate before the steady-state formation is established i.e., during the initial transient from take-off (from the ground) to the achievement of the final optimal formation for the long flight. Some arguments can be offered for stability when communication aspects are introduced during bird flight. The model developed in this article does not consider the energy expenditure during bird rotation, or energy expenditure during honking. Honking for geese could be an additional way of losing heat to the environment [10]. Birds communicate with sound, and possibly also react to signals received by their surface sensors for temperature and pressure, and possibly from other signals like odor and humidity.

When the flock exchanges information e.g., during the change of the leader, or a call (honk) for a bird to not fall behind, this information must be transmitted to the entire flock as a disturbance that does not cause instability for the formation. To maintain formation, a perturbation in the position should not amplify and lead to a pattern collapse. In the case of the "V" formation, a pattern collapse clearly will not lead to a chaotic state, with undefined entropy generation, but rather to a state that is defined in the worst case by the individual bird flight i.e., Case 1 discussed above. However, for such a change there is an expected rise in the bird temperature. We can argue that when the pattern is one where the maximum entropy is already being generated (i.e., the system is at an extremum), it will be less likely that a short burst low amplitude perturbation will magnify into a major disruptive amplification or create a pattern-destroying condition especially if the initial perturbation is small. Some further clues for understanding how communication ties into entropy generation and stability are available from connections between the entropy of information sciences and entropy generation [43-48]. Regardless, it should be noted that we have not examined the implications of communication for stability-enhancement in any rigorous manner in this article. Honking does not appear to impact the stability of the formation. When operating at a maximum rate of entropy production a temporary increase in entropy could perhaps be easily damped out. A stable steady state is considered stable if *any* arbitrarily small or large perturbation of that state can only decrease (dampen) with time. To test for stability, perturbation of *all* the state variables must be studied which is not possible without accurate temperature and pressure measurements during flight at all chosen altitudes. The second differential of the equation $\beta$ with x/L, almost approaches zero, in units of inverse Kelvin ($K^{-1}$) i.e., therefore is likely an inflection point. Effectively, this means that even a large perturbation in power (a possibility from a loud honk) may not be able to easily alter the existing entropy generation rate without an easily perceived change in the environment, thus giving the birds ample time to fall back into formation as they can quickly respond to a temperature variation.

Imagine a perturbation in the energy expenditure of the lead bird that crosses the cell boundaries. In the model described in this article, this perturbation must travel to the rest of the flock. For a "V" formation it can effectively travel across both legs of the "V" because of the symmetry. A perturbation that can impact every control volume can thus impact the microstates across the system (flock). If the rate of entropy generation per unit of power expended ($\beta$) is already maximized *and* the entropy generate rate is already maximized, there is no further possibility of instability becoming magnified. The magnitude of the entropy generation or even a negentropy change from the information exchanges may also not compare with the entropy generation from the flight itself, especially for large-bird flocks, thus preventing instability.

Although it is not clear how much mass is associated with the information transfer with sound-based bird-to-bird communication, a comparison of entropy rate perturbations can be made in principle with Equation (C1c) or similar for other flock formations if the sound duration is known. The Landauer principle [43] is a physical principle for assessing the lower theoretical limit of energy consumption during information exchanges. The principle posits that the irreversible manipulation of information, such as the erasure of a bit or the merging of two computation paths, should be accompanied by a corresponding entropy increase in non-information-bearing degrees of freedom of the information-processing apparatus or its environment [43-47]. Thus, information transfer energy can lead to instabilities that can amplify. A gain in entropy can potentially cause a loss in information as well as cause instability. Any information that has a physical representation, must be embedded in the statistical mechanical degrees of freedom of a physical system [35-39]. Brillouin [46] and Szilard [47] inferred that changing an information-bit requires at least a kT(ln (2)) amount of energy. Correspondingly k(ln (2)) is a minimum entropy perturbation for a small amount of information (a bit) (k is the Boltzmann constant). This is a small quantity unless multiplied by the Avogadro number. Regardless, chaotic results can arise even with small energy disturbances [22]. The stability of the avian pattern needs more study than is offered in this article. The "V" formation permits minor adjustments to the pattern to accommodate slightly different size birds by providing adequate freedom of lateral positioning.

In addition, there are self-synchronization aspects that should be considered that are related to wing movement that can impact stability or cause instability from wingtip eddies. For heavy aircraft, such wingtip eddies cause instabilities that can be problematic to other aircraft that follow their flight path. The syncing of oscillatory agents can generate dynamics for large amplitude perturbations that can resonate with the environment, presumably also the basis of vortex enhanced energy benefits [10]. It is possible that the amplification can also lead to destructive amplification of perturbations when the energy source is active. An example of destructive amplification is the well-known London-bridge condition. Here the amplification of perturbations with human lock-step synchronization led to a near-collapse of the bridge. There are alternate ways of examining such instabilities [49]. From Figure 2, we note that the birds are only somewhat synchronized as far as their wing positions are concerned, suggesting that there is *no* tendency for lock-step wing self-synchronization that could sometimes become the magnifier of an instability nucleating perturbation. The results presented in Reference 10, appear to also indicate that some synchronization may lead to beneficial energy savings. Because wing tip eddies could be destructive, there is a need to study this aspect of eddy-related interaction between birds in greater detail.

Self-organization is known to invoke features of interaction that are internal to the control volume with events happening at a smaller scale than the overall pattern dimensions [4,50,51]. Note that the *sweep area* provided by the birds (shown in Figure 1) is also minimized for the V formation flights. The differences in the entropy generation rate will magnify as the *sweep area* falls. It is known that the metabolic rate of birds decreases with body mass and not only with the environment temperature [10,40]. New pathways for defects (debris) to be ejected are also possibly created for efficiently self-organized bird formations. For the bird flight conditions that involve many birds in a formation, these could involve new pathways for the removal of waste or the *creation of additional wingtip eddies*. These have not been considered in the simple model presented in this article. The energy savings (for enhancing the endurance) could also increase if the loss in the mass of the birds from converting food to waste during flight is considered.

With a simple thermal model, we have inferred that the SV formation compared to other formations is the ideal pattern for optimal energy rate expenditure at least during long flights. The thermodynamic model presented herein is suggestive of rapid thermal responses in birds being critical for formation flying. The SV and other closely related formations allow spatial separation that allows the optimal bird temperature during the flight passage. This formation also ensures that all trailing birds have the same thermal environment. As the flock size increases, the power expended per bird falls, when the comparison is made for the same velocity or mass flow rate. Compared to when the bird flies by herself, formation-flying in a flock of birds thus reduces fatigue for each bird, leading to larger distances flown by the flock.

# 5. Conclusions

A thermal model for an avian flight indicates that:

(i) Formation flying leads to optimization of the energy used.
(ii) Formation flying permits possible flights at higher altitudes.
(iii) As the flock size increases the energy efficiency and the rate of entropy generation increase.
(iv) The entropy generation rate is higher with the SV formation compared with the SVT organization.
(v) No difference is noted between the ST and SV formations for the rate of entropy generation.

The model validates three main experimental observations, namely, (i) the preference for a V formation over other formations, (ii) the energy-savings benefit of flying the V formation, and (iii) the decrease in energy employed as the flock size increases. The MEPR postulate for avian flight-formation predicts the observed formation, namely the "V" or SV formation. We conclude that for the long-distance flights by large-size migratory birds, the flock of birds is arranged in their most optimal energy-efficient *and* stable pattern configuration.

The analysis indicates that the organization of live objects may follow the MEPR principle like that noted previously in chemical and metallurgical processes that display self-organization behavior [4,51].

Finally, it must be pointed out that this is only one model for avian formation flying. There are other equally persuasive published models [1,2,3,6,9]. Unless detailed and accurate thermal and pressure gradient measurements become available, it is not possible to be confident about the assumptions made in this article for the thermodynamic model. It is also possible that a hybrid model with eddies *and* thermals may be required for comprehensively comparing the entropy generation rate between various formations (patterns).

**Funding:** This research received no external funding
**Conflicts of Interest:** No conflict of interest.

# 6. References

[1] Mirzaeinia A., Heppner F., Hassanalian M., An analytical study on leader and follower switching in V-shaped Canada Goose flocks for energy management purposes, Swarm Intelligence (2020) 14:117–141, https://doi.org/10.1007/s11721-020-00179-x

[2] Norberg U. M. L., Flight and scaling of flyers in nature, WIT Transactions on State of the Art in Science and Engineering, Vol 3, © 2006 WIT Press doi:10.2495/1-84564-001-2/2d

[3] Rayner J.M., Estimating power curves of flying vertebrates, J Exp Biol. 1999 Dec;202(Pt 23):3449-61.

[4] Sekhar, J.A., Self-Organization, Entropy Generation Rate, and Boundary Defects: A Control Volume Approach. Entropy **2021**, 23, 1092. https://doi.org/10.3390/e23081092

[5] S. G. Tzafestas, Energy, Information, Feedback, Adaptation, and Self-organization, Intelligent Systems, Control and Automation: Science and Engineering 90, 2018 https://doi.org/10.1007/978-3-319-66999-1_9, Springer International Publishing AG

[6] Meir J. U, et.al., Reduced metabolism supports hypoxic flight in the high-flying bar-headed goose (Anser indicus), https://elifesciences.org/articles/44986

[7] Reis A H, Use and validity of principles of extremum of entropy production in the study of complex systems, Annals of Physics, 2014, Vol. 346, July Pages 22-27

[8] Martyushev L. M., Maximum entropy production principle: history and status" *Phys. Usp.* 2021, 64 (6)

[9] Portugal, S., Hubel, T., Fritz, J. *et al.* Upwash exploitation and downwash avoidance by flap phasing in ibis formation flight. *Nature* **505,** 399–402 (2014)., https://doi.org/10.1038/nature12939

[10] Torre-Bueno J.R., Temperature regulation and heat dissipation during flight in birds, J Exp Biol, 1976 Oct;65(2):471-82., PMID: 1003090

[11]https://ocw.mit.edu/ans7870/16/16.unified/propulsionS04/UnifiedPropulsion4/UnifiedPropulsion4.htm Last accessed December 21, 2021

[12] Sekhar J.A., The description of morphologically stable regimes for steady-state solidification based on the maximum entropy production rate postulate, J. Mater. Sci., Vol. 46, 6172-6190, 2011.

[13] Veveakis, E.; Regenauer-Lieb, K. Review of extremum postulates. Curr. Opin. Chem. Eng. 2015, 7, 40–46
[14] Lucia U., Maximum entropy generation and K-exponential model, Physica A: 2010, Statistical Mechanics and its Applications, vol. 389, pp 4558–4563
[15] Hill A., Entropy production as the selection rule between different growth morphologies. Nature, 348**,** 426–428, 1990. https://doi.org/10.1038/348426a0
[16] Martyushev L.M, Seleznev V.D., and Kuznetsova I. E., Application of the Principle of Maximum Entropy production to the analysis of the morphological stability of a growing crystal, Zh. Éksp. Teor. Fiz. 200, 118 pp. 149
[17] Ziman J.M., The general variational principle of transport theory, Can. J. Phys. 1956 35 pp. 1256
[18] Kirkaldy J. S., Entropy criteria applied to pattern selection in systems with free boundaries, Metall. Trans. A. 16A, pp. 1781-1796, 1985.
[19] Bensah, Y. D., Li, H. P., & Sekhar, J. A. (2012). The Sgen Rate Maximization Postulate: Applications to Process-Path Analysis for Solidification and Micropyretic Synthesis. In Key Engineering Materials (Vol. 521, pp. 79–86). Trans Tech Publications, Ltd., https://doi.org/10.4028/www.scientific.net/kem.521.79
[20] Ziegler H., and Wehrli C., On a principle of maximal rate of entropy production, J. Non-Equilib. Therm., 12 pp. 229, 1978.
[21] Cavagna A., Irene Giardina,Francesco Ginelli, Thierry Mora, Duccio Piovani,, Raffaele Tavarone, and Aleksandra M. Walczak, Dynamical maximum entropy approach to flocking, Physical Review, 2014, E **89**, 042707
[22] Latora V., Baranger M., Rapisarda A, Tsallis C., The rate of entropy increase at the edge of chaos, Physics Letters A, 2000, Volume 273, Issues 1–2 Pages 97-103
[23] Prigogine I, Introduction to Thermodynamics of Irreversible Processes, 3rd ed., Interscience Publ., NY, 1967.
[24]. Ziegler H, An Introduction to Thermomechanics, North-Holland Publ. Co., Amsterdam, 1977
[25] Nive R. K., Simultaneous Extrema in the Entropy Production for Steady-State Fluid Flow in Parallel Pipes https://arxiv.org/abs/0911.5014],
[26] Matyushev L. M., and Selzenev V. D., Maximum entropy production principle in physics, chemistry and biology, Physics Reports **426**, 1-45 (2006)]
[27] Endres R.G., Entropy production selects nonequilibrium states in multistable systems. Sci Rep **7,** 14437 (2017). https://doi.org/10.1038/s41598-017-14485-8
[28] Kim E.-j., Information Geometry, Fluctuations, Non-Equilibrium Thermodynamics, and Geodesics in Complex Systems. *Entropy* **2021**, *23*, 1393. https://doi.org/10.3390/e23111393
[29] Martyushev L.M., Nazarova A.S., Seleznev V.D., J Phys A, 2007, 40:371
[30] Klaidon Axel, Naturwissenschaften, 2009,96:653
[31 Onsager L., Reciprocal relations in irreversible processes I, Phys. Rev. **37**, 405-426 (1931).
[32] https://en.wikipedia.org/wiki/Stationary-action_principle. Last accessed November 1, 2021
[33] Feynman R.P., 1985 (seventh printing, 1988), QED: The Strange Theory of Light and Matter, Princeton University Press, ISBN 0-691-02417-0, pp. 51–2.
[34] https://www.britannica.com/science/thermoreception/Birds. Accessed November 10, 2021
[35] Cavina, E., The "Organ of flight": Paratympanic Organ (PTO) of Vitali in Wild Birds as Biological Barometer-Altimeter. *Academia Letters*, Article 1613, 2021. https://doi.org/10.20935/AL1613.

[36] Schwab, R.G. and Schafer, V.t F., "AVIAN THERMOREGULATION AND ITS SIGNIFICANCE IN STARLING CONTROL" (1972). Proceedings of the 5th Vertebrate Pest Conference (1972). 25. https://digitalcommons.unl.edu/vpc5/25
[37] Smit et al., Climate Change Responses, 2016 3:9 DOI 10.1186/s40665-016-0023-2
[38] http://people.eku.edu/ritchisong/birdmetabolism.html Last Accessed December 22, 2021
[39] McNabb F. M. A., , Critical Reviews in Toxicology, 2007, 37:163–193
[40] Rezende E. L., Pez-calleja M. V., and Bozinovic F., July 2001, The Auk, 118 (3):781–785,
[41] Veghte, J. H. and Herried, C. F. , Radiometric determination of feather insulation and metabolism of arctic birds. Physiol. Zool. 1965, 38, 367-75.
[42] Ehrlich P. R. Dobkin D. S. and Wheye D., in https://web.stanford.edu/group/stanfordbirds/text/essays/Temperature_Regulation.html. Copyright 1988 by Paul R. Ehrlich, David S. Dobkin, and Darryl Wheye. Last Accessed December 24, 2021.
[43] Landauer R., (1961), Irreversibility and heat generation in the computing process", IBM Journal of Research and Development, 5 (3): 183–191, doi:10.1147/rd.53.0183, retrieved 2015-02-18. Cited from https://en.wikipedia.org/wiki/Landauer_principle, Last accessed November 1, 2021
[44] https://en.wikipedia.org/wiki/Entropy_in_thermodynamics_and_information theory. Last accessed November 1, 2021
[45] Natal, J.; Ávila, I.; Tsukahara, V.B.; Pinheiro, M., Maciel, C.D. Entropy: From Thermodynamics to Information Processing. Entropy **2021**, 23, 1340. https://doi.org/ 10.3390/e23101340
[46] Brillouin L., The Negentropy Principle of Information, Journal of Applied Physics 24, 1152 (1953); https://doi.org/10.1063/1.1721463
[47] Szilard L., Über die Ausdehnung der phänomenologischen, Thermodynamik auf die Schwankungserscheinungen, Zeitschrift für Physik, 1925, 32: 753-788. DOI:10.1007/BF01331713, Corpus ID: 121162622
[48] Dawson, D.A.; Sid-Ali, A.; Zhao, Y.Q. Local Stability of McKean–Vlasov Equations Arising from Heterogeneous Gibbs Systems Using Limit of Relative Entropies. *Entropy* **2021**, *23*, 1407. https://doi.org/10.3390/e23111407
[49] Joshi V, Srinivasan M., Walking crowds on a shaky surface: stable walkers discover Millennium Bridge oscillations with and without pedestrian synchrony. Biol. Lett., 2018 14: 20180564. http://dx.doi.org/10.1098/rsbl.2018.0564
[50] Vanchurin V, Wolf Y. I., Katsnelson M. I., and Koonin E.V. PNAS 2022 Vol. 119 No. 6 e2120037119,
[51] Yaw B. D. and Sekhar J. A., Solidification Morphology and Bifurcation Predictions with the Maximum Entropy Production Rate Model, **Entropy 2020**, vol. 22, doi:10.3390/e22010040